\documentclass{ws-procs9x6}            
\graphicspath{{Figs/}}
\begin{document}

\newcommand{\boldsigma}{{\bm \sigma}}
\newcommand{\be}{\begin{equation}}
\newcommand{\ee}{\end{equation}}
\newcommand{\rf}[1]{~(\ref{#1})}
\newcommand{\bea}{\begin{eqnarray}}
\newcommand{\eea}{\end{eqnarray}}
\newcommand{\bd}{\begin{displaymath}}
\newcommand{\ed}{\end{displaymath}}
\newcommand{\ba}{\begin{array}}
\newcommand{\ea}{\end{array}}
\newcommand{\bc}{\begin{center}}
\newcommand{\ec}{\end{center}}
\newcommand{\bfl}{\begin{flushleft}}
\newcommand{\efl}{\end{flushleft}}
\newcommand{\bfr}{\begin{flushright}}
\newcommand{\efr}{\end{flushright}}
\newcommand{\f}{\frac}
\newcommand{\capitolul}{\chapter}
\newcommand{\au}{\u{a}}
\newcommand{\ai}{\^{a}}
\newcommand{\ii}{\^{\i}}
\newcommand{\s}{\c{s}}
\newcommand{\tc}{\c{t}}
\newcommand{\un}{\underline}
\newcommand{\xiq}{\xi({\bf q})}
\newcommand{\vs}{\vspace}
\newcommand{\bQ}{{\bf Q}}
\newcommand{\bq}{{\bf q}}
\newcommand{\bk}{{\bf k}}
\newcommand{\bB}{{\bf B}}
\newcommand{\bx}{{\bf x}}
\newcommand{\bz}{{\bf z}}

\def\csch{\mathop{{\rm csch}}}
\def\sech{\mathop{{\rm sech}}}
\def\ua{\uparrow}
\def\da{\downarrow}
\def\ket#1{\left\vert #1 \right\rangle}
\def\dg{\dagger}


\def\br{{\bf r}}\def\bR{{\bf R}}\def\bs{{\vec }}
\def\bk{{\bf k}} \def\bK{{\bf K}}\def\bq{{\bf q}} \def\bp{{\bf p}}
\def\bQ{{\bf Q}}

\def\dwave{$d_{x^2-y^2}$-wave~}
\def\da{\downarrow} \def\ua{\uparrow} \def\la{\leftarrow}
\def\ra{\rightarrow}

\def\al{\alpha} \def\be{\beta} \def\si{\sigma}
\def\ga{\gamma} \def\de{\delta} \def\ve{\varepsilon} \def\ep{\epsilon}
\def\ze{\zeta} \def\et{\eta} \def\th{\theta}
\def\vt{\vartheta} \def\ka{\kappa} \def\lam{\lambda}
\def\ss{\sigma} \def\ta{\tau}
\def\ph{\phi} \def\vp{\varphi}
\def\om{\omega} \def\Ga{\Gamma} \def\De{\Delta}
\def\Th{\Theta} \def\La{\Lambda} \def\Si{\Sigma} 
\def\Ph{\Phi} \def\Ps{\Psi} \def\Om{\Omega}

\def\sm{\small} \def\la{\large} \def\La{\Large}
\def\LA{\LARGE} \def\hu{\huge} \def\Hu{\Huge}
\def\ti{\tilde} \def\wti{\widetilde}
\def\non{\nonumber}
\def\bra{\langle}
\def\ket{\rangle}
\def\={\!\!\!&=&\!\!\!}
\def\+{\!\!\!&&\!\!\!+~}
\def\-{\!\!\!&&\!\!\!-~}
\def\id{\!\!\!&\equiv&\!\!\!}


\def\CC{CeCoIn$_5$}
\def\CI{CeIrIn$_5$}
\def\CS{CeCu$_2$Si$_2$}

\def\UR{URu$_2$Si$_2$}
\def\UP{UPd$_2$Al$_3$}

\def\CB{CeB$_6$}
\def\YB{YbB$_{12}$}

\def\CF{CeFeAsO$_{1-x}$F$_x$}
\def\LF{LaFeAsO$_{1-x}$F$_x$}

\def\bk{{\bf k}}
\def\bq{{\bf q}}
\def\bQ{{\bf Q}}


\title{Resonant spin excitations in unconventional heavy fermion superconductors and Kondo lattice compounds}

\author{Peter Thalmeier$^*$}

\address{Max Planck Institute for Chemical Physics of Solids,\\
D-01187 Dresden, Germany \\
$^*$E-mail: thalm@cpfs.mpg.de\\
http://www.cpfs.mpg.de}

\author{Alireza Akbari}
\address{Asia Pacific Center for Theoretical Physics,\\
Dept. of Physics, and Max Planck POSTECH Center for Complex Phase Materials, 
POSTECH, Pohang 790-784, Korea\\
E-mail: alireza@apctp.org\\
and\\
Max Planck Institute for Solid State Research,\\
D-70569 Stuttgart, Germany\\
}

\begin{abstract}
The heavy quasiparticle bands in Kondo materials which originate in the hybridization of f- and conduction electrons exhibit numerous, sometimes coexisting, broken symmetry phases. Most notable are unconventional superconductivity, itinerant small moment antiferromagnetism and hidden order of higher order multipoles of f-electrons which all lead to a gapping of the heavy bands. In rare cases the chemical potential lies within the hybridization gap and the ground state is a Kondo semiconductor without ordering.
The dynamical magnetic response of such gapped f-electron systems has been investigated with inelastic neutron scattering. It was found that collective spin exciton modes which are due to residual quasiparticle interactions appear below the threshold of superconducting or hidden order gap or directly the hybridzation gap . The spin exciton resonance is commonly located around a zone boundary vector \bQ~ with nesting properties in the normal state. In the superconducting case its appearance gives a strong criterion for the gap symmetry requesting a sign change $\Delta_{\bk+\bQ}=-\Delta_\bk$ due to the coherence factors.
Therefore this many body effect with fundamental importance 
may also be used as a tool to discriminate between proposed gap models. While the spin resonance has been observed for many compounds we restrict our discussion here exclusively to the small group of  f-electron superconductors  \CC, \CS~and  \UP , hidden order Kondo compounds \CB~and \UR~as well as the Kondo semiconductor \YB.
\end{abstract}

\keywords{Heavy fermion superconductors, Kondo lattice, feedback effect, spin exciton}

\bodymatter

\section{Introduction}
\label{sec:intro}

The f-electron based heavy fermion systems, mostly intermetallic Ce- and U- compounds, are schematically described by Anderson-lattice type models which contain conduction electrons, localized f-electrons and a hybridization term. In mean-field approximation the effect of on-site f-electron correlations is taken into account by imposing a total on-site charge constraint on the average. Then the appearance of heavy hybridized bands and associated narrow density of states (DOS) peak around the Fermi level with a width of the order of the Kondo temperature $T_K$ may be naturally explained \cite{newns:87}. They are responsible for the typical heavy fermion anomalies in thermodynamic and transport quantities \cite{hewson:93}.

Furthermore residual interactions of the heavy quasiparticles which are on-site repulsive but inter-site attractive may lead to the formation of unconventional Cooper pairs and associated superconducting (SC) gap function \cite{miyake:86,thalmeier:05}. Its nontrivial symmetry implies the existence of nodes where the gap vanishes, leading to a power law behavior of the density of states (DOS) of SC quasiparticle excitations.
The latter show up as power laws in the temperature dependence of physical quantities like specific heat, thermal conductivity and NMR relaxation which are the first typical signature of an unconventional superconductor. The unconventional gap symmetry not only 
modifies the low energy quasiparticle spectrum as compared to the fully gapped s-wave case.  In favorable situation it may also lead to the appearance of new collective triplet spin exciton magnetic modes inside the SC gap by the feedback effect. The corresponding quasiparticle bound state poles in the dynamical magnetic response function lie at resonance frequencies below the two quasiparticle creation threshold. The wave vector of collective modes is usually close to a nesting vector of the normal state. The observation of a spin resonance is directly tied to the nodal structure of the gap and may be used as a tool to constrain the latter. In this way unexpectedly inelastic neutron scattering (INS) has turned out to be an important tool to investigate non-trivial gap structure of heavy-fermion superconductors. 

In this work we review in some detail the theoretical spin exciton models for superconducting \CC, \CS~and \UP.  
The large wave vector collective spin exciton is a bound state of quasiparticles. Its physical nature is therefore different from the small wave vector collective "Higgs'-like modes that are associated with long-wavelength amplitude fluctuations of the superfluid density which are directly connected with the spontaneous symmetry breaking \cite{pekker:15}. The latter can usually be observed only indirectly through coupling to other lattice modes.

Surprisingly spin exciton-like modes have also been found in non-superconducting heavy fermion systems. This is possible because the bound state poles may appear whenever there is a near-singular frequency dependence of the real part of the bare susceptibility close to a gap threshold. In the SC feedback system this is ensured by the unconventional coherence factors. However in the case of normal state hybridization gaps or hidden order related gaps it will always be true. This is the reason why the spin exciton was found in the small hybridization gap Kondo semiconductor \YB~and the hidden order gapped compounds \UR~and \CB~which will all be discussed in detail in this review.

As a final case the rather special superconducting Ce-4f electron based pnictide \CF~will be presented. It exhibits an indirect spin exciton effect which is due to the electronic two -component structure consisting of localized 4f electrons and itinerant 3d electrons in different layers of the crystal. The SC spin exciton which appears in the 3d layers is coupled weakly to the 4f excitations in the Ce layer producing a peculiar feedback effect which is related to the one in \UP~with its dual 5f electron system.

\section{Electronic Structure of Kondo lattice compounds}
\label{sec:electronic}


The quasiparticles in heavy electron systems may be schematically obtained as low energy 
excitations of the Anderson lattice model. As a minimum ingredient it contains the conduction electron
dispersion $\ve_\bk$ (due to a hopping energy t) , localized and degenerate $(m=1...N_f)$ f-electron states $\ve^f_{\bk m} = \ve^f_\bk$ with strong on-site Coulomb interaction $U_{ff}$ and, most importantly a hybridization $V_{\bk m}=V_\bk$ between localized and conduction states. The model may be extended by adding crystalline electric field (CEF), orbital dependent hybridization and Zeeman splitting of the f-level. Furthermore spontaneous order of quasiparticles such as superconductivity, magnetism or hidden order may be described by adding pairing or molecular field terms to the Hamiltonian. First we discuss only the minimum $SU(N_f)$ model and the various additional terms will be added when required. It is defined by
\bea
{\cal H}
&=&
\sum\limits_{{\bf k}m }
\Bigl[\varepsilon^c_{\bf k} c_{{\bf k}m}^{\dagger}c_{{\bf k}m}+
\varepsilon^f_{\bf k} f_{{\bf k}m} ^{\dagger}f_{{\bf k}m}
+V_{{\bf k}}\left( c_{{\bf k}m}^{\dagger}f_{{\bf k}m}
+h.c.\right)\Bigr] \nonumber \\
&&+\sum\limits_{i,m\neq n} U_{ff}f_{im}^{\dagger}f_{in}f_{in}^{\dagger}f_{im}.
\label{eq:HA}
\eea
Here $m=(\tau,\sigma)$ is generally comprised of orbital pseudo spin ($\tau$) and Kramers pseudo spin ($\sigma $) degrees of freedom of the CEF ground state multiplet.
The large on-site Coulomb energy $U_{ff}\gg t,|\ve_f|$ leads to strong correlation effects, i.e. strong reduction of f-electron double occupancies.  In the limit $U_{ff}\ra\infty$ they are forbidden and this limit  may be described by introducing an auxiliary slave boson field $b_i$ for the f holes supplemented  by the local constraint that  the total on-site charge $Q_i=n_{fi}+n_{bi}$ is equal to one where $n_{fi}=\sum_m f^\dag_{mi}f_{mi}$  and $n_{bi} =b^\dag _{i}b_{i}$ \cite{coleman:84}. On the lattice this constraint may only be implemented in mean field approximation assuming $\bra b_i\ket=b$ everywhere \cite{newns:87}. Then the above Hamiltonian reduces to an effective one body Hamiltonian
\begin{equation}
{\cal H}
=
\sum\limits_{{\bf k}m}
\varepsilon ^c_{{\bf k}}c_{{\bf k}m}^{\dagger}c_{{\bf k}m }+
\tilde{\varepsilon}^f_{\bf k} f_{{\bf k}m} ^{\dagger}f_{{\bf k}m}
+\tilde{V}_{{\bf k}}\left( c_{{\bf k}m }^{\dagger}f_{{\bf k}m}
+h.c.\right) 
+\lambda(r^2-1).
 \label{eq:HAMF}
\end{equation}
Here $\lam$ is a Lagrange parameter introduced to enforce the charge constraint.
It shifts the effective f-electron level $ \tilde{\ve}^f_{\bk m}$ close to the chemical potential.  
The effective hybridization $\tilde{V}_{{\bf k}}$ is renormalized by the slave boson mean field amplitude $\bra b_i\ket = r$.
Together we have
\bea
\tilde{V}^2_\bk = r^2V^2_\bk = V^2_\bk(1-n_f)
;\;\;\;
\tilde{\ve}^f_{\bk m} &=& \ve^f_{\bk m}+\lam.
\eea
In our subsequent discussion of magnetic response we assume a simplified form of the Anderson model that ignores
the \bk~ dependence (but not necessarily orbital dependence) of hybridization. This means we effectively set $\tilde{V}_\bk = \tilde{V}$.
For our purpose this simplification is sufficient, but it cannot always be used, e.g. for the discussion of electronic structure in the pseudogap Kondo insulator CeNiSn \cite{ikeda:96} or the topological insulators like SmB$_6$ \cite{hanzawa:98,takimoto:11,dzero:10}.
\begin{figure} 
\hspace{0.5cm}
\includegraphics[width=0.90\textwidth]{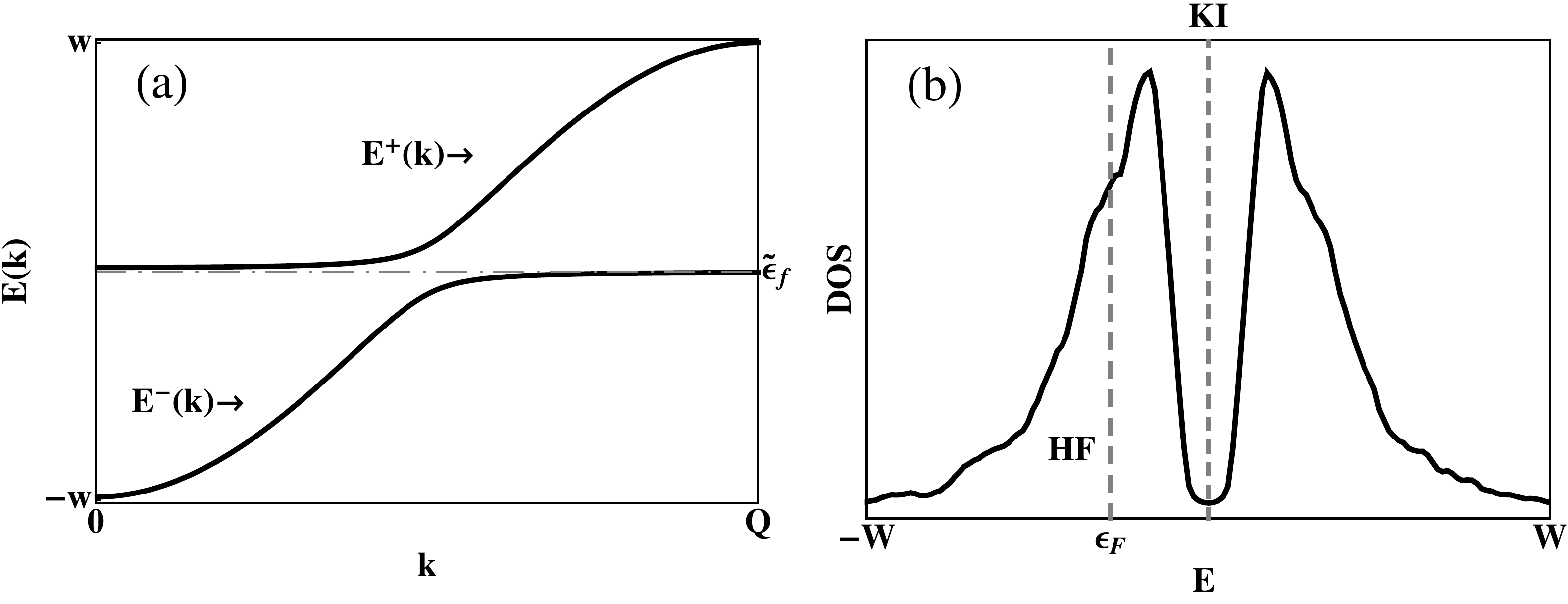}
\caption{(a) The hybridized quasiparticle bands around the renormalized f-level $\tilde{\epsilon}_f$  and (b) corresponding DOS for band parameters of typical hybridized bands. Fermi level position for heavy fermion metal (HF, like \CB) and Kondo insulator (KI, like \YB) is indicated. For KI it is inside the hybridization charge gap $\Delta_c$.} 
\label{fig:Quasiband} 
\end{figure} 
The single particle type mean field Hamiltonian may be diagonalized leading to a quasiparticle Hamiltonian
\bea
{\cal H}_{MF}&=& \sum\limits_{i,{\bf k},m,\alpha}\ve^{\alpha}_{{\bf k}}
a^{\dagger}_{\alpha,{\bf k}m}a_{\alpha,{\bf k}m}+\lambda(r^2-1),\non\\
\ve^{\pm} _{\bf  k}&=&\frac{1}{2}
\bigl[\epsilon^{c}_{{\bf  k}}+\tilde{\epsilon}^{f}_{{\bf  k}}\pm\sqrt{(\epsilon^{c}_{{\bf  k}}
-\tilde{\epsilon}^{f}_{{\bf  k}})^2+4\tilde{V}^2_{{\bf  k}}}\bigr]
\label{eq:HMFqp}
\eea
where $\ve^{\pm} _{\bf  k}$ 
are the pair ($\alpha =\pm$) of hybridized quasiparticle ($a_{\alpha{\bf k}m}$) bands, each $N_f$-fold degenerate. The indirect 
gap in Fig.~\ref{fig:Quasiband} is of the order of the Kondo temperature: $\ve^+_0 -\ve^-_\bQ\simeq T_K$ with $T_K=W\exp(-1/(JN_c(0))$ for $N_f=2$. ($W, N_c(0)$ are conduction band width and DOS, respectively, $J=2V^2/|\ve_f|$ is the effective on-site exchange constant.)
The unitary transformation to these states is given by 
\bea
f_{{\bf k}m }=u_{+, {\bf k}} a_{+,{\bf k}m }+u_{-, {\bf k}} a_{-,{\bf k}m };\;\;\;
c_{{\bf k}m }=u_{-, {\bf k}} a_{+,{\bf k}m }-u_{+, {\bf k}} a_{-,{\bf k}m }.
\eea
with the admixture coefficients defined by 
\bea
2u_{\pm, {\bf k}}^2 =
1\pm (\epsilon^{c}_{{\bf  k}}-\tilde{\epsilon}^{f}_{{\bf  k}})/\sqrt{(\epsilon^{c}_{{\bf  k}}
-\tilde{\epsilon}^{f}_{{\bf  k}})^2+4\tilde{V}^2_{{\bf  k}}}.
\eea
They appear as matrix elements in the numerator of the expression for the bare magnetic response functions,
possibly together with coherence factors of the condensed (superconducting or other) phase.

\section{Feedback effect of gap formation in the superconducting phase on dynamic magnetic response}
\label{sec:feedback}

The mean field slave boson theory leads to non-interacting hybridized heavy quasiparticles as sketched in Fig.~\ref{fig:Quasiband}. However, inclusion of fluctuations beyond mean field approximation \cite{doniach:87,riseborough:92} introduces residual scattering between the quasiparticles. This has twofold consequences. Firstly it enhances the spin fluctuations which may mediate an unconventional superconducting state \cite{miyake:86,thalmeier:05}. Secondly once the gap is established a superconducting feedback effect may modify the bare spin response strongly at the gap threshold. In conjunction with the quasiparticle interactions this leads to the appearance of a new collective magnetic mode inside the superconducting gap, commonly called spin-exciton bound state or resonance. In the heavy fermion superconductors \CC, \CS~and \UP~where it has been found it  may be described within a phenomenological treatment containing  the heavy quasiparticle dispersion $\ve^\al_\bk$ $(\al=\pm$), their interaction $J_\bq$ and the unconventional gap function $\Delta_\bk$. While the former two may be reasonably well modeled the latter is usually unknown or several candidates may exist. Then the investigation of the predicted collective modes may give important information on the nodal structure of the possible gap functions. Thus the superconducting feedback effect is not only an interesting many-body effect in itself but also provides important clues on the symmetry of the unconventional gap function.

\begin{figure} 
\hspace{1.9cm}
\includegraphics[width=0.6\textwidth]{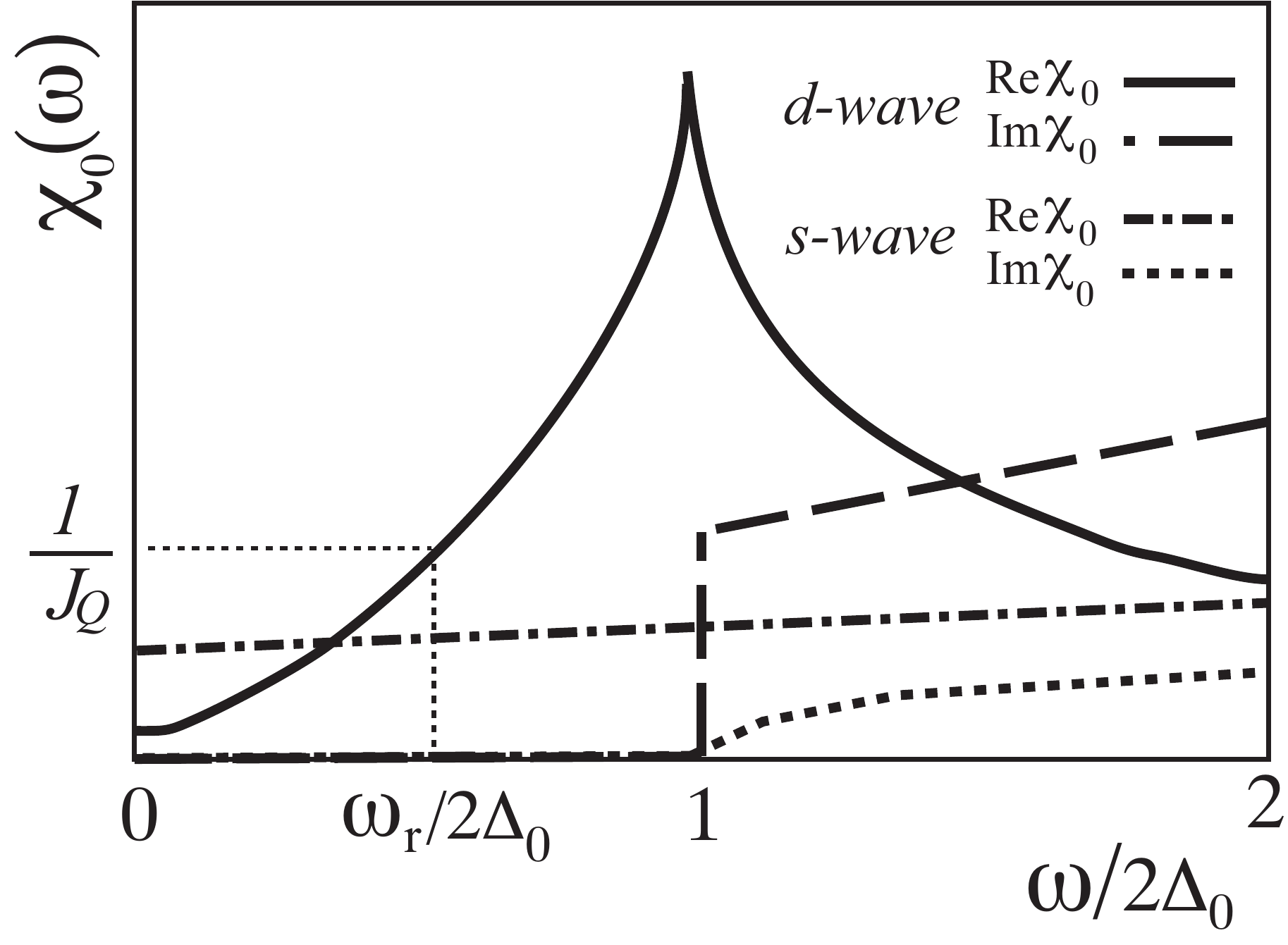}
\caption{Schematic frequency dependence of bare susceptibility $\chi_0(\bQ,\om)$ (real and imaginary parts) for gap functions without (s-wave) and with (d-wave) sign change $\Delta_{\bk+\bQ}=\pm\Delta_\bk$, respectively. The singular behavior of   $Re\chi_0(\bQ,\omega)$  leads to the formation of a spin exciton bound state  according to Eq.~(\ref{eq:rescond}) at wave vector \bQ~ and energy $\omega_r(\bQ)<2\Delta_0$.} 
\label{fig:Feedback} 
\end{figure}

First we discuss the feedback effect in the superconducting state. For clarity here we suppress the indices $(\al,m)$ of hybridized bands and omit their matrix elements. Then the bare magnetic response, restricting to contributions from creation of two quasiparticles \cite{norman:00} ($T\ll\De$) at wave vector \bQ~  is given by
\bea
\chi_0({\bf Q},\omega) &=&\sum_{\bf k}
F({\bf k},{\bf Q})\frac{f(E_{\bk+\bQ})+f(E_\bk)-1}
{\omega-(E_{{\bf k}+{\bf Q}}+E_{\bf k})+i\eta}\non\\
F({\bf k},{\bf Q})&=&\frac{1}{4}\Bigl[
1-\frac{\epsilon_{{\bf k}}\epsilon_{{\bf k}+{\bf Q}}
+\Delta_{{\bf k}}\Delta_{{\bf k}+{\bf Q}}}
{E_{{\bf k}}E_{{\bf k}+{\bf Q}}}\Bigr]
\label{eq:Lindhard}
\eea
where $E_\bk=[\ve^2_\bk+\Delta_\bk^2]^\frac{1}{2}$ and $F({\bf k},{\bf Q})$ are superconducting quasiparticle energy and coherence factor, respectively. The latter exhibits two principally different behaviors at gap threshold $\Omega_c=min_{\bk \in FS}(|\De_\bk|+|\De_{\bk+\bQ}|)\approx 2\De_0$ for creation of two quasiparticles out of the condensate: 
When the gap function has no sign change at \bQ, meaning $\De_{\bk+\bQ}=\De_\bk$ ,e.g. for an s-wave gap, then $F(\bk,\bQ)\rightarrow 0$ vanishes, leading to a soft onset of $Im\chi_0(\bQ,\omega)$ above $\omega\approx 2\Delta_0$ and a corresponding smooth behavior
of  $Re\chi_0(\bQ,\omega)$ around the threshold. On the other hand, for an unconventional nodal gap with  $\De_{\bk+\bQ}=-\De_\bk$,
 $F(\bk,\bQ)\rightarrow \frac{1}{2}$ leading to a steplike increase of $Im\chi_0(\bQ,\omega)$ above $\omega\approx 2\Delta_0$ and a corresponding singular peak in  $Re\chi_0(\bQ,\omega)$. These two distinct cases are sketched in Fig.~\ref{fig:Feedback}.

The effective non-retarded quasiparticle interaction $J_\bq$ will enhance the dynamical magnetic response at certain favorite wave vectors, in particular at those connected with the static nesting properties of the heavy electrons. Within RPA approximation the spectrum of spin fluctuations at general wave vector \bq~ is given by
\bea
Im \chi_{RPA}({\bf q},\omega)=
\frac{Im \chi_0({\bf q},\omega)}
{(1-J_{\bf q}Re\chi_0({\bf q},\omega))^2+J_{\bf q}^2(Im\chi_0({\bf q},\omega))^2}\non\\
\label{eq:imchi}
\eea
which is directly proportional to the dynamical structure function and cross section in INS. 
In the normal state this simplifies to a low frequency AF paramagnon type expression 
$Im \chi_{RPA}({\bf Q},\omega)\approx\chi_{RPA}({\bf Q})\om\om_{sf}/(\om^2_{sf}+\om^2)$ which peaks at the paramagnon (spin fluctuation) energy scale $\om_{sf}(T)$. The interaction of quasiparticles caused by exchanging these overdamped bosonic spin fluctuations leads to unconventional superconductivity in the first place \cite{monthoux:92}. Once the gap is established 
it is obvious that $Im \chi_{RPA}({\bf q},\omega)$ in Eq.~(\ref{eq:imchi}) should be very sensitive around $\bq\approx\bQ$ to the absence or presence of the sign change $\De_{\bk+\bQ}=\pm\De_\bk$. In the latter case the first part in the denominator of $Im\chi_{RPA}({\bf q},\omega)$ may vanish leading to a bound state pole or resonance (for finite but small $Im\chi_0({\bf q},\omega)$). For $T\ll T_c$ its resonance position $\omega_r(\bQ)$ is given  by the condition 
\bea
\frac{1}{J_{\bf Q}}=\sum_{\bf k}\frac{F({\bf k},{\bf Q})}
{\omega_r(\bQ)-(E_{{\bf k}+{\bf Q}}+E_{\bf k})}
\label{eq:rescond}
\eea
illustrated also by the graphical solution in Fig.~\ref{fig:Feedback}. Since the singular  $Re\chi_0(\bQ,\omega)$ enabling the solution is only present close $\bq\approx\bQ$ where the sign change appears, the resonance is usually confined to the vicinity of this wave vector.
Obviously $\omega_r(\bQ)/2\Delta_0<1$ must be fulfilled for a true spin exciton resonance split from the continuum of SC quasiparticle excitations. In the known cases (Table \ref{table:resonance}) this is indeed the case when $\omega_r$ from INS is compared to the value of $2\Delta_0$ from tunneling experiments. The spin exciton existence and dispersion $\omega_r(\bq)$ for general $\bq\approx\bQ$ is determined by two effects: i) the increase in $\Omega_c$ due to the fact that the sign change condition for the gap is no longer exactly valid  ii) the reduction of the quasiparticle interaction which may be modeled by $J_\bq=J_\bQ\Gamma^2_\bQ[(\bq-\bQ)^2+\Gamma^2_\bQ]^{-1}$. Both effects support confinement of the bound state solution close to $\bq\approx\bQ$.

In the above discussion it is assumed that the superconducting gap is much smaller than the normal state hybridization gap. The latter is the indirect gap in Fig.~\ref{fig:Quasiband}a,b which is of the order of the Kondo energy scale $T_K$. Therefore this condition is fulfilled for T$_c \ll$ T$_K$ which is usually the case in heavy fermion superconductors. The physical reason for the appearance of the superconducting feedback resonance may be summarized in a concise way: The sign change of an unconventional gap function for translation by wave vector \bQ~ leads to a constant coherence factor  $F(\bk,\bQ)\rightarrow \frac{1}{2}$ for small energies. This means that for sign-changing gap the bare magnetic response is that of a 'semiconductor' at the gap threshold $\sim 2\Delta_0$, therefore the excitonic states (in the magnetic spin triplet channel) may exist. This leads to the expectation that the latter are not necessarily tied to the superconducting state. One might suspect that the spin exciton can appear inside the hybridization gap itself or other type of gaps even without superconductivity. Such cases of hybridized f-electron compounds have indeed been found as discussed in Sec.~\ref{sec:nonSC}.
\begin{figure} 
\hspace{.25cm}
\includegraphics[width=0.46\textwidth]{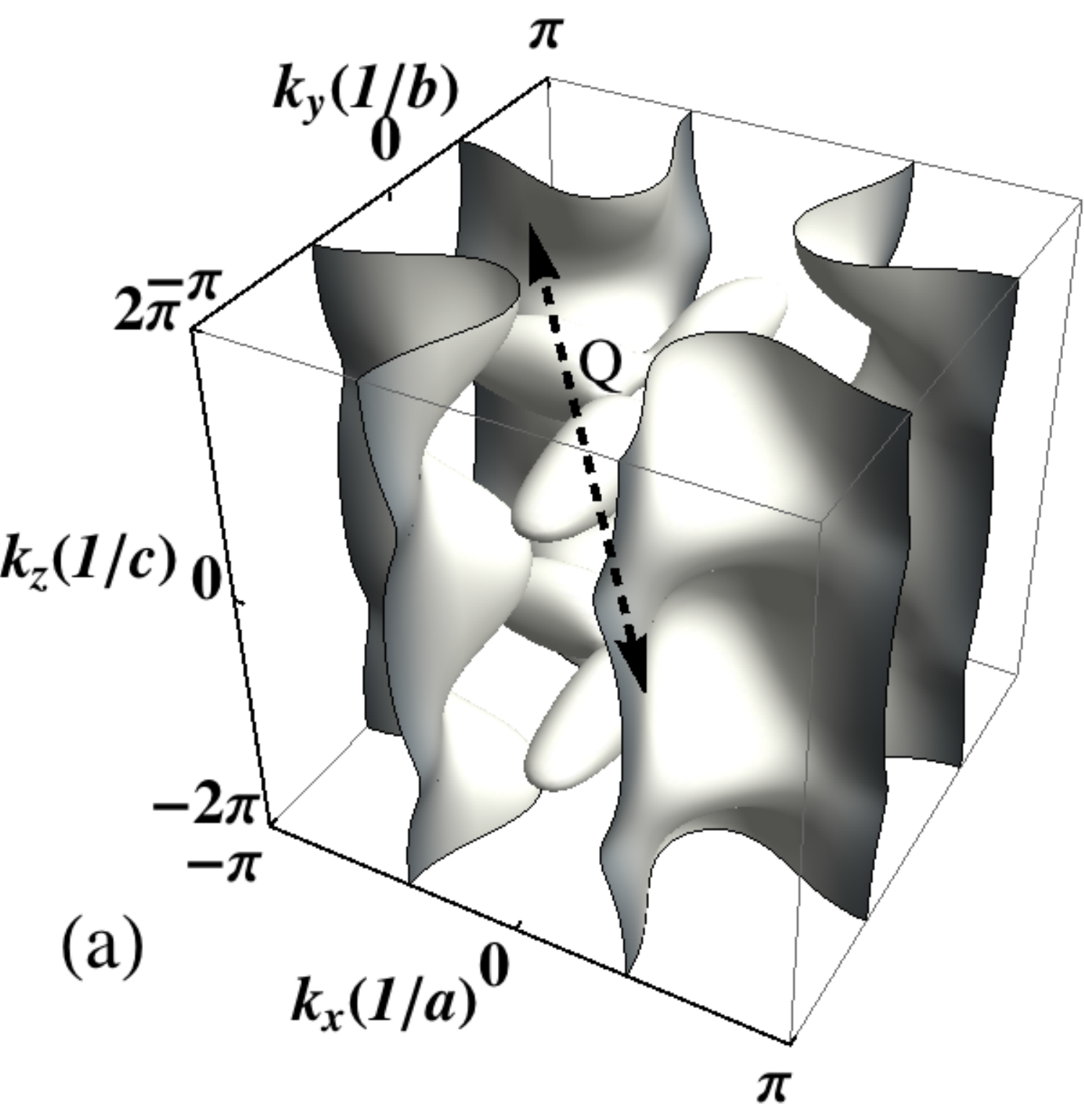}
\hspace{.5cm}
\includegraphics[width=0.4\textwidth]{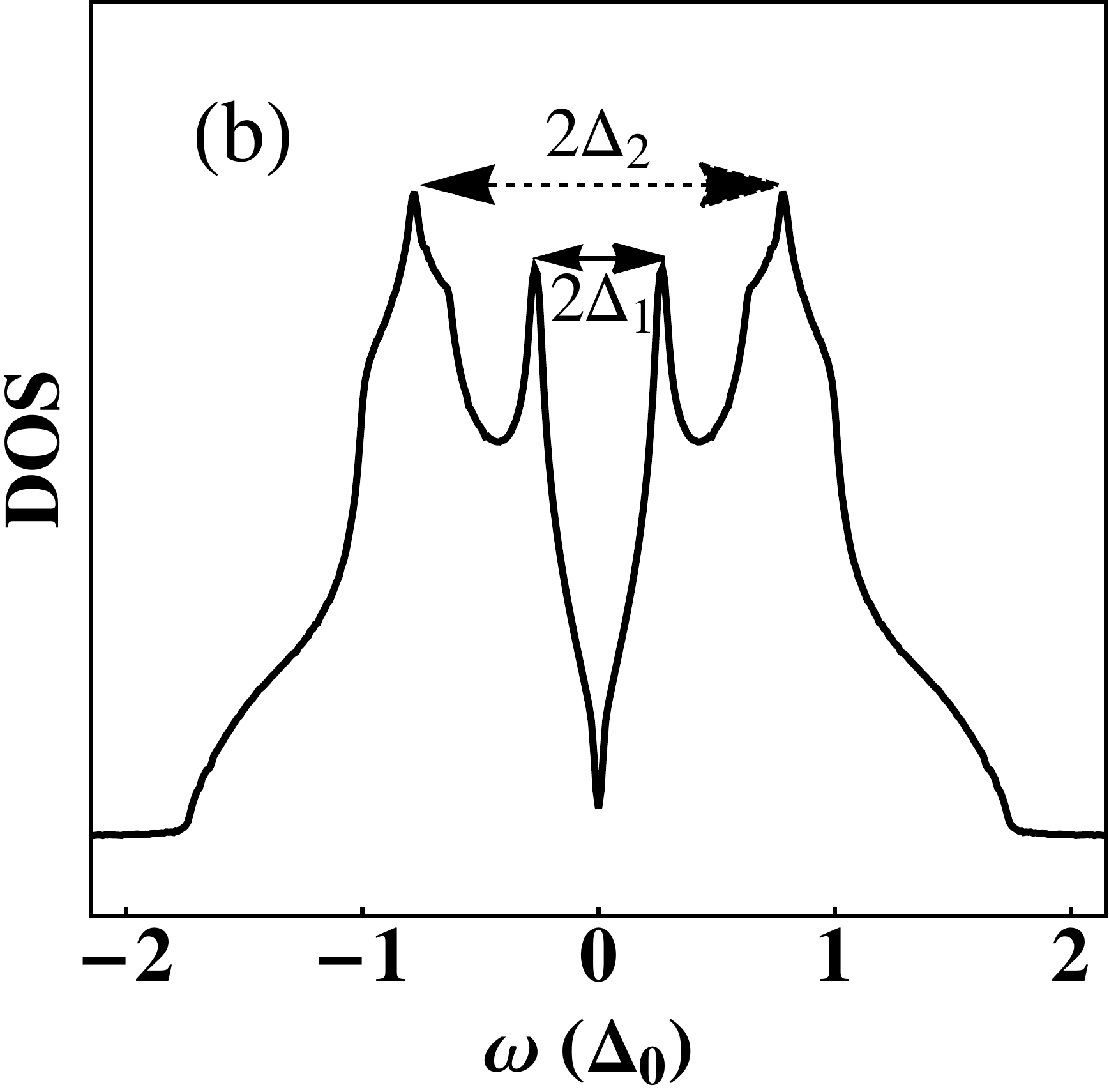}
\caption{(a) The Fermi surface of \CC~obtained from mean field hybridized quasiparticle band structure \cite{akbari:11}.
Nesting vector $\bQ=(\frac{1}{2},\frac{1}{2},\frac{1}{2})$ (r.l.u.) between different M$(\pi,\pi)$ point columns is indicated.
(b) Quasiparticle DOS (unit: $1/\Delta_0$) of \CC~in the SC $d_{x^2-y^2}$ state with $\Delta_\bk=\frac{1}{2}\Delta_0(\cos k_x-\cos k_y)$. Here the main gaps on the center and corner FS sheets in (a) are $2\Delta_1=0.56\Delta_0$ and   $2\Delta_2=1.5\Delta_0$.} 
\label{fig:Fermi_B0} 
\end{figure}

\section{Spin excitons in unconventional superconductors}
\label{sec:USC}

The previous discussion suggests that spin excitons should appear naturally in unconventional superconductors.
In fact numerous examples have been observed in the cuprates \cite{eschrig:06} and pnictides \cite{inosov:10,korshunov:08}  where the d-wave or s$_{\pm}$ type gap functions, respectively,  exhibit a sign change under the in-plane translation vector $\bQ =(\frac{1}{2},\frac{1}{2},0)$ of the tetragonal BZ. Well defined spin excitons with energies below $2\Delta_0$ are found at this wave vector. For cuprates they show a typical 'hourglass' dispersion around \bQ. However, it should be mentioned that the latter has also been identified in non-superconducting cuprates and cobaltates  where they have been attributed to dynamic stripe formation or spiral magnetic correlations (e.g. Ref.~\citenum{drees:13}). In the unconventional heavy fermion superconductors these possibilities are excluded. There are three clear examples of spin exciton modes  that have been observed at much lower energies ( $\sim 1$ meV range) which we will now discuss in greater detail.

\subsection{Field dependent spin exciton in \CC~}
\label{subsec:115}

The CeMIn$_5$ intermetallic compounds (M= Co, Ir, Rh) \cite{onuki:04} are model heavy fermion systems which show
both antiferromagnetism (AF) and superconductivity as function of substitution at M sites. The latter appears commonly
around the AF quantum critical point which may also be tuned by hydrostatic pressure or external magnetic field. 
Recently these phases have also been investigated in artificial quasi-2D superlattice structures \cite{mizukami:11} with alternating magnetic Ce and nonmagnetic Yb layers.

Superconductivity is most prominent in \CC~with the highest $T_c=2.3$ K reported in heavy fermion systems. The order parameter symmetry has been among the most controversial with singlet $d_{x^2-y^2}$ (node lines along 
diagonals of the BZ)  or $d_{xy}$ (node lines parallel to BZ axes) as originally obtained from field angle-resolved thermal conductivity \cite{izawa:01} and specific heat \cite{aoki:04} measurements, respectively. The observation of a pronounced spin exciton resonance \cite{stock:08} and its theoretical analysis \cite{eremin:08} gave a strong support for the  $d_{x^2-y^2}$ model. 
Subsequently this has also been confirmed by specific heat investigations at lower temperatures \cite{an:10}. A more direct approach to this symmetry issue is provided by quasiparticle interference (QPI) spectroscopy which was proposed to give a fingerprint of the nodal orientation \cite{akbari:11}. This has  been confirmed in experiments \cite{allan:13,zhou:13} that directly obtain the angular (\bk-) dependence of the gap function confirming the essential  $d_{x^2-y^2}$ character.
Therefore \CC~is a prime example where the observation of a spin resonance has given first important clues on the symmetry of the gap function. For magnetically isotropic case the triplet spin exciton should split into three modes in a magnetic field. The splitting, though of different type,  has been found for the first time for the resonance mode in \CC~  \cite{stock:12} and was explained within RPA model calculations  \cite{akbari:12a}.

\begin{figure} 
\includegraphics[width=0.51\textwidth]{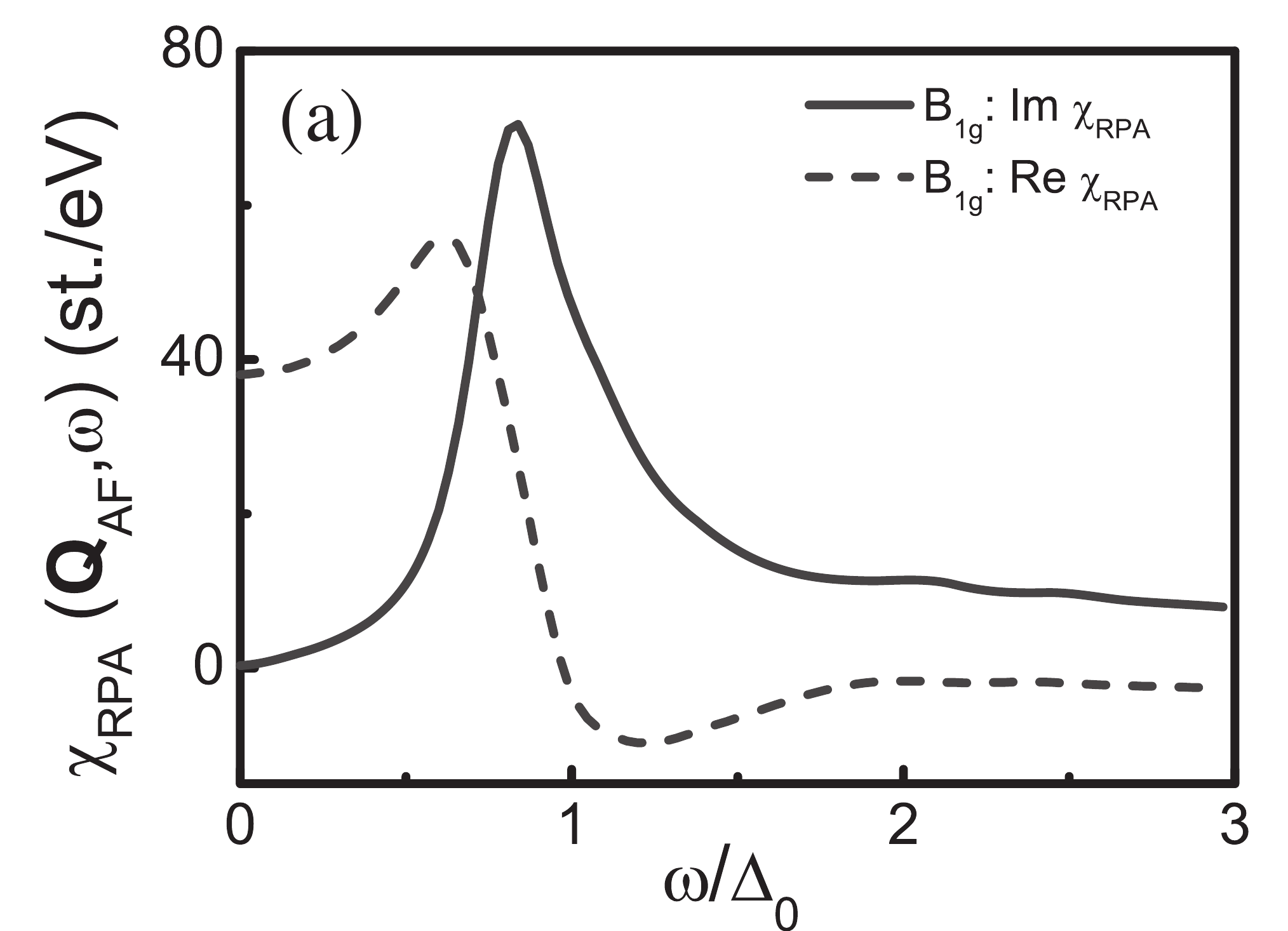}\hspace{-.02cm}
\includegraphics[width=0.48\textwidth]{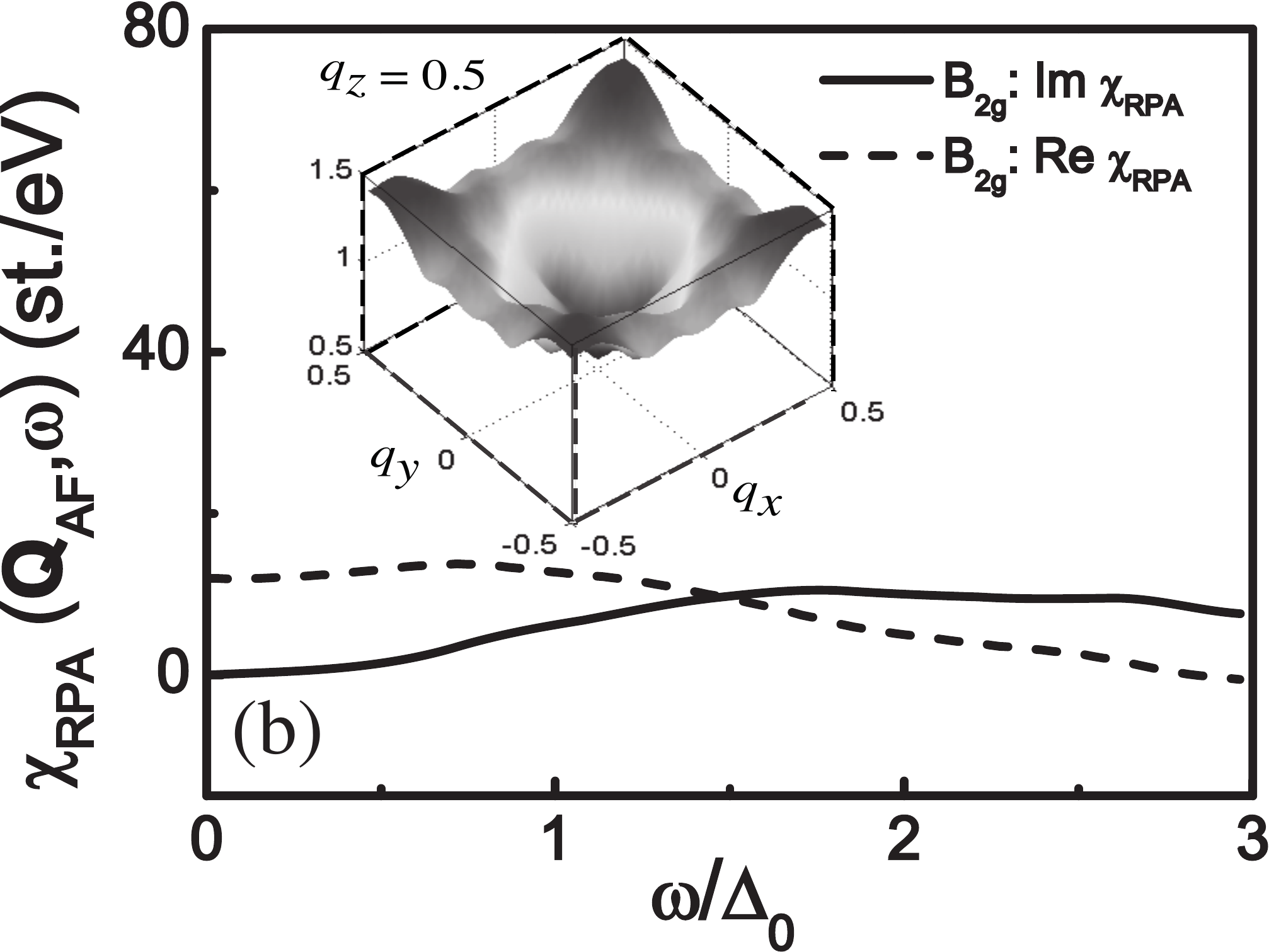}
\caption{RPA spectral function (imaginary part) and corresponding real part for \CC~in (a) the $B_{1g}$ $(d_{x^2-y^2})$ and (b) the $B_{2g}$ $(d_{xy})$ superconducting state. The inset shows the static Lindhard function $Re\chi_0(\bq,0)$ in the ($q_x,q_y,0.5$) (r.l.u.) plane containing the zone boundary nesting vector \bQ~ of the maximum (Ref.~\citenum{eremin:08}).} 
\label{fig:CeCoIn5_res} 
\end{figure} 

As in all other cases a theoretical analysis requires a reasonably realistic model for the underlying heavy quasiparticle bands. Local density approximation (LDA) band structure calculations  \cite{shishido:02,maehira:03} lead to a multisheet compensated Fermi surface structure where the band14  electron sheet has the largest volume. The LDA band mass which does not incorporate the strong electron correlations is too small by about a factor 20. The mass renormalisation can be achieved within the phenomenological mean-field Anderson lattice type model described in Sec.\ref{sec:electronic}. This was carried out for the above main sheet in Ref.~\citenum{tanaka:06} by fitting the \bk- dependence of unhybridized bands and the hybridization matrix elements such that the band14 surface is reproduced. The resulting model Fermi surface is shown in Fig.~\ref{fig:Fermi_B0}a and it qualitatively agrees with the LDA result, including the proper nesting vector $\bQ=(\frac{1}{2},\frac{1}{2},\frac{1}{2})$ (r.l.u.) between the corrugated M$(\pi,\pi)$ - point columns. This leads to  a pronounced maximum of the the static normal state Lindhard function $\chi(\bq)$ at \bQ~ shown in the inset of   Fig.~\ref{fig:CeCoIn5_res}b.
\begin{center}
\begin{table}
\tbl{Compilation of experimental spin resonance characteristics in superconducting (top part) and non-superconducting (bottom part) heavy fermion metals and Kondo insulators. Here $2\Delta_0$ and $\Delta_c$ denote the quasiparticle charge gap in the SC and non-SC cases, respectively. The resonance appears around nesting  vector \bQ.}
{\begin{tabular}{lrrrrrr}
Compound & $T_c$       & $2\Delta_0$     & $\omega_r$ &  $\frac{\omega_r}{2\Delta_0}$ & \bQ       &Ref.\\
                   &[K]             &[meV]              & [meV]           &                                                  & [r.l.u.]   &    \\
\hline\hline
CeCu$_2$Si$_2$ & \hspace{0.0cm} 0.60 & \hspace{0.0cm} 0.26 & \hspace{0.0cm} 0.20 & \hspace{0.0cm}0.78 &  \hspace{0.0cm} (0.215,0.215,1.458)  & \cite{stockert:08,stockert:11} \\
CeCoIn$_5$ & 2.30 & 0.92 & 0.60  & 0.65 & (0.5,0.5,0.5)  & \cite{stock:08,stock:12}\\
UPd$_2$Al$_3$ & 1.80 & 0.86 & 0.35 & 0.40& (0.,0.,0.5) & \cite{sato:01,hiess:06}\\
\hline
   & $T_{HO}$       & $\Delta_c$     & $\omega_r$ &  $\frac{\omega_r}{\Delta_c}$ &        & \\
\hline
URu$_2$Si$_2$ & 17.8& 4.1& 1.86  & 0.45&  (0.,0.,1.)     & \cite{bourdarot:10,aynajian:10}   \\
CeB$_{6}$  & 3.2 & 1.3  & 0.5  &  0.39  &  (0.5,0.5,0.5)   &  \cite{friemel:12, paulus:85}  \\
YbB$_{12}$ & - & 15 & 15  & $\sim$ 1&  (0.5,0.5,0.5)     & \cite{nemkovski:07,okamura:05}   \\
\hline
\end{tabular}}
\protect\label{table:resonance}
\end{table}
\end{center}
In the superconducting state the order parameter has a twofold effect through gap formation and modification of dipole matrix elements between the SC quasiparticle eigenstates as expressed by the  coherence factors. 
For the  $d_{x^2-y^2}$ (B$_{1g}$) gap function $\De(\bk)=(\Delta_0/2)(\cos k_x-\cos k_y)$ one has a {\it different} sign of the gap on FS sheets in opposite M - point corners of Fig.~\ref{fig:Fermi_B0}a connected by \bQ. This leads, via the feedback mechanism, to a pronounced resonance peak in the dynamic magnetic response (Fig.~\ref{fig:CeCoIn5_res}a) at this wave vector. On the other hand for the  $d_{xy}$ (B$_{1g}$) gap function $\De(\bk)=(\Delta_0/2)\sin k_x\sin k_y$ the node lines are parallel to the axes, this means that FS sheets in opposite corners of the BZ have the {\it same} sign of the gap function. Consequently no magnetic resonance peak at the connecting wave vector \bQ~ should be expected, as is demonstrated by  Fig.~\ref{fig:CeCoIn5_res}b. When a similar calculation is
performed for gap functions belonging to the remaining tetragonal representations \cite{eremin:08} no clear resonance peak is found.
Therefore the clearcut observation of the latter in INS experiments \cite{stock:08} strongly supported the  $d_{x^2-y^2}$ (B$_{1g}$) type gap function for \CC~and thus resolved the issue of gap function symmetry in favor of the original proposal made from thermal conductivity in rotating fields \cite{izawa:01}. Later the  $d_{x^2-y^2}$ symmetry has been directly confirmed by investigation of QPI spectra in \CC~ \cite{akbari:11,allan:13,zhou:13}. Furthermore thermal magnetotransport \cite{kasahara:08} experiments have suggested that this is also the proper SC gap symmetry for the \CI~compound. It would therefore be interesting to check with INS whether \CI~exhibits  a similar spin exciton mode in the SC state.\\
\begin{figure} 
\includegraphics[width=0.42\textwidth]{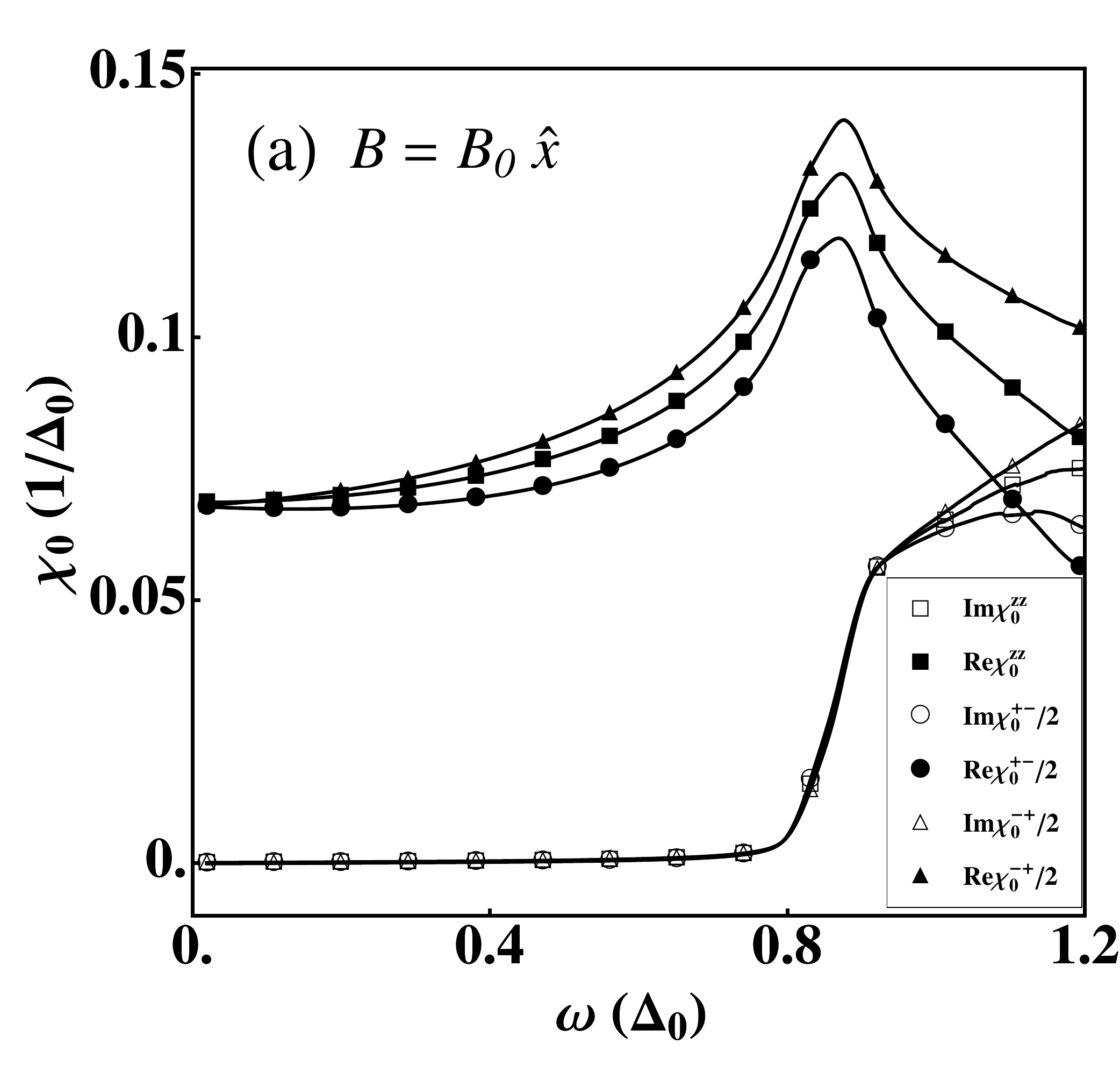}\hfill
\includegraphics[width=0.55\textwidth]{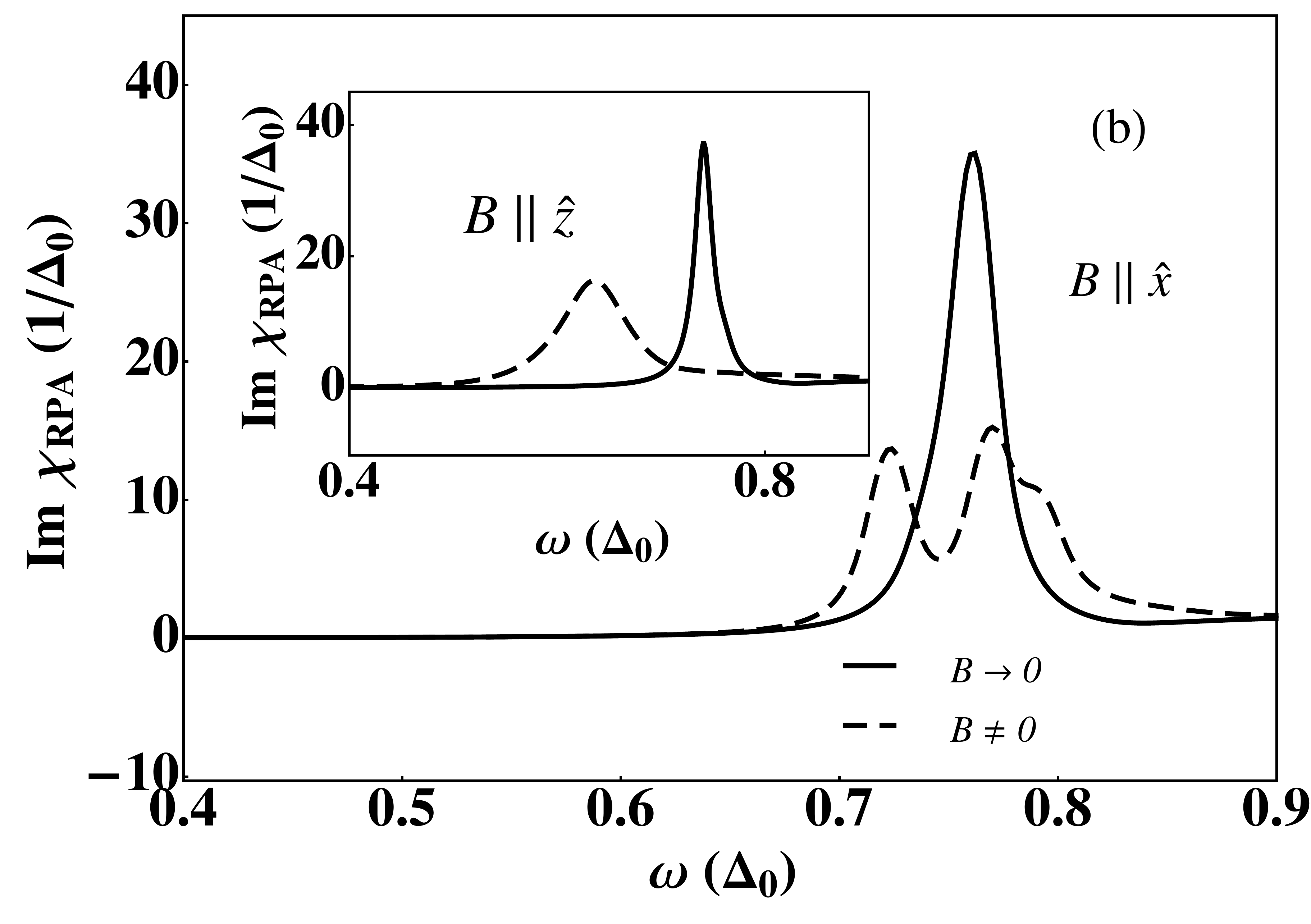}
\caption{(a) Real and imaginary part of bare pseudo spin susceptibility in magnetic field  $h_B^f=\frac{1}{2}g_{\alpha}^{f}\tilde{\chi}\mu_BB =0.3\Delta_0$ ($g_\alpha$ are effective g-factors of ground state doublet with $\alpha =\parallel,\perp$ with respect to tetragonal plane;  here $g^f_\perp/g^f_\parallel = m_\perp/m_\parallel = 2.3$; $\tilde{\chi}$ is the Stoner enhancement factor). The peak height of the real part is different for the three components which leads to the splitting of the collective mode from the RPA susceptibility shown in (b) for field direction $\bB=\mbox{B}\hat{\bx}\; (\parallel a)$. For $\bB=\mbox{B}\hat{\bz}\; (\parallel c)$ (inset) no splitting but only a downward shift and broadening occur (Ref~\citenum{akbari:12a}).} 
\label{fig:splitspec} 
\end{figure} 

The spin exciton is a triplet collective mode, then one may expect that the application of a magnetic field in any direction should  lead to a splitting into three components. This was indeed predicted \cite{ismer:07} in context of the cuprates, although experimentally not confirmed so far. Evidence of the splitting was, however, reported for 11- type Fe-pnictide superconductors \cite{bao:10}. The similar experiment for \CC~ \cite{stock:12,raymond:12} gave the first evidence for spin exciton splitting in heavy fermion superconductors, albeit with a surprising modification: i) The splitting is only observed for tetragonal in-plane field, for field along the c-axis only a broadening and downward shift of the resonance occurs. ii) For field in the plane the resonance splits into only {\it two} instead of the expected three  longitudinal and transverse polarized modes. To explain these observations it is not possible to argue on the level of Zeeman splitting of localized CEF states, but rather one has to investigate how the full many-body RPA response of heavy quasiparticles is modified in the presence of a Zeeman term and CEF introduced effective anisotropies:
The heavy bands will split in a magnetic field leading to different longitudinal and transverse bare Lindhard functions which changes the resonance condition in the denominator of the RPA dynamic susceptibility. This may ensue the splitting or broadening of the collective spin exciton depending on field direction  \cite{akbari:12a}.
The latter enters through the CEF potential that imprints an effective g-factor anisotropy on the $\Gamma_7^{(1)}$ ground state doublet, in addition the effective quasiparticle interaction $\hat{J}_\bq$ will be an anisotropic (uniaxial) tensor. This leads to a collective RPA susceptibility that has both out-of plane (longitudinal zz) and in plane (circularly porlarized $\pm, \mp$) contributions which add differently for in-plane ($\bB=\mbox{B}\hat{\bx}$) and out-of plane ($\bB=\mbox{B}\hat{\bz}$) fields. We obtain \cite{akbari:12a}
\begin{figure} 
\hspace{2.59cm}
\includegraphics[width=0.55\textwidth]{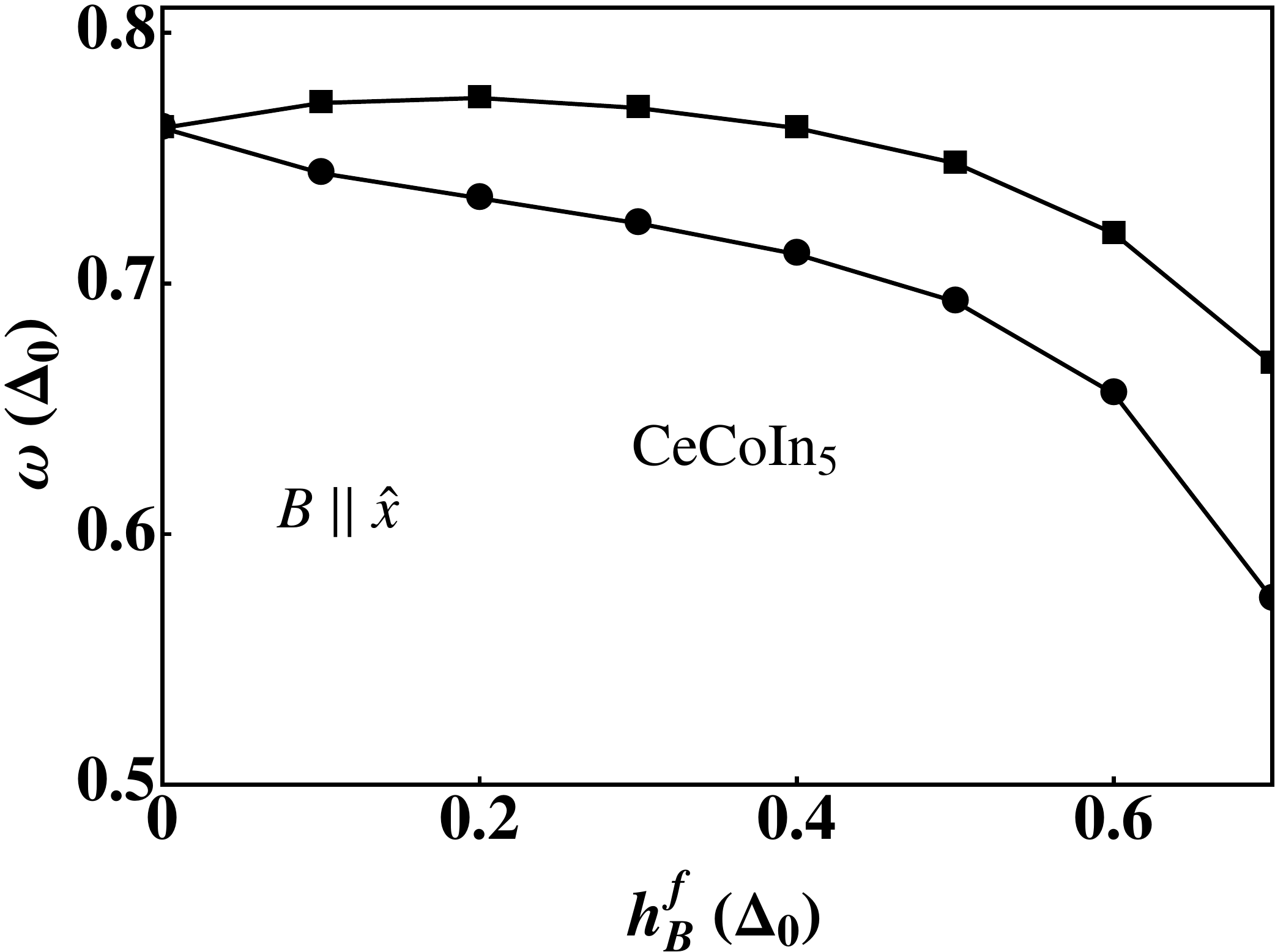}
\caption{Split doublet peak positions of  spin exciton resonance as function of dimensionless field strength  $h_B^f/\Delta_0$
for $B\parallel x$. Almost symmetric linear small field splitting becomes strongly nonlinear for larger fields (Ref~\citenum{akbari:12a}).} 
\label{fig:splitting_field} 
\end{figure} 
%
\bea
\mbox{B}\hat{\bx}: && \chi_{RPA}({\bf q},\omega)= 
 \frac{m_{\parallel}^2\chi^{zz}_{0}}{1-\lambda_{\parallel}\chi^{zz}_{0}}\nonumber\\
 && +\frac{(m_{\parallel}^2+m_{\bot}^2) (\chi^{+-}_{0}+\chi^{-+}_{0})-(\lambda_{\parallel}m_{\bot}^2+\lambda_{\bot}m_{\parallel}^2)\chi^{+-}_{0}\chi^{-+}_{0} }
  {4-(\lambda_{\parallel}+\lambda_{\bot})(\chi^{+-}_{0}+\chi^{-+}_{0})+\lambda_{\bot}\lambda_{\parallel}\chi^{+-}_{0}\chi^{-+}_{0}},
 \nonumber\\ [0.3cm]
\mbox{B}\hat{\bz}: &&\chi_{RPA}({\bf q},\omega)=
 \frac{m_{\perp }^2\chi^{zz}_{0}}{1-\lambda_{\perp}\chi^{zz}_{0}}
+  \frac{m_{\parallel }^2\chi^{+-}_{0}}{2-\lambda_{\parallel}\chi^{+-}_{0}} + \frac{m_{\parallel }^2\chi^{-+}_{0}}{2-\lambda_{\parallel}\chi^{-+}_{0}}.
 \label{eq:chiRPA}
\eea
where the combined interaction parameters are defined by $\lambda_l=m_l^2J_\bq^l$ ($l=\parallel,\perp$ to tetragonal plane) and $m_l, J^l_\bq$ are anisotropic CEF ground state matrix elements and quasiparticle interactions, respectively.
Due to the sign change $\Delta_{{\bf k}+{\bf Q}}=-\Delta_{{\bf k}}$ for nesting  momentum  ${\bf Q}, $ Im$\chi_0(\bQ,\omega)$ remains  zero for the low frequencies and then shows  a sudden jump at the onset frequency of the quasiparticle continuum  close to $\Omega_c={\rm min}(|\Delta_{{\bf k}}|+|\Delta_{{\bf k}+{\bf Q}}|)$. This happens around $2\Delta_1\simeq\Delta_0$ , where $2\Delta_1$ is the main gap in the SC DOS in (Fig.\ref{fig:Fermi_B0}.b) obtained from the tunneling spectrum of \CC~\cite{rourke:05}.
 The resonance may appear for energies  $\omega<\Omega_c$,  provided that   (i) $J^{ll'}_{{\bf q}}{\rm Re} \chi_{0{\bf q}}^{ll'}(\omega)=1$ and (ii)  ${\rm Im} \chi_{0{\bf q}}^{ll'}(\omega)\simeq 0$ ($ll' = +-, -+, zz$).
For $m_\parallel^2J_{\bf q}^\perp \approx m_\perp^2J_{\bf q}^\parallel$ it is possible that for $B \parallel x$ the resonance condition is fulfilled only for $\chi_0^\pm, \chi_0^\mp$ but not for $\chi_0^{zz}$ leading to a split doublet resonance. This behavior is indeed observed in the main Fig.~\ref{fig:splitspec}a,b. On the other hand for $B \parallel z $ no splitting but only a broadening and downward shift occurs as shown in the inset of Fig.~\ref{fig:splitspec}b. For $\bB=\mbox{B}\hat{\bx}$ the splitting of the bare susceptibility peak in Fig.~\ref{fig:splitspec}a is symmetric, however the  field dependent splitting of the resonance positions in Fig.~\ref{fig:splitspec}b show asymmetry already for moderate fields. This is due to the fact that they are determined by the zeroes in the {\it denominators} in the RPA susceptibility of Eq.~(\ref{eq:chiRPA}) and therefore the splitting must be asymmetric even though that of the  $\chi_0^\pm, \chi_0^\mp$ peaks is symmetric. For higher fields a large nonlinear downward shift  occurs in Fig.~\ref{fig:splitting_field} before the resonance peaks can no longer be identified. It has been proposed that a spin exciton condensation into a SDW phase inside the SC phase occurs at large fields \cite{michal:11}.

\subsection{\CS: spin exciton at incommensurate wave vector}
\label{subsec:122}

\begin{figure} 
\hspace{.15cm}
\includegraphics[width=0.45\textwidth]{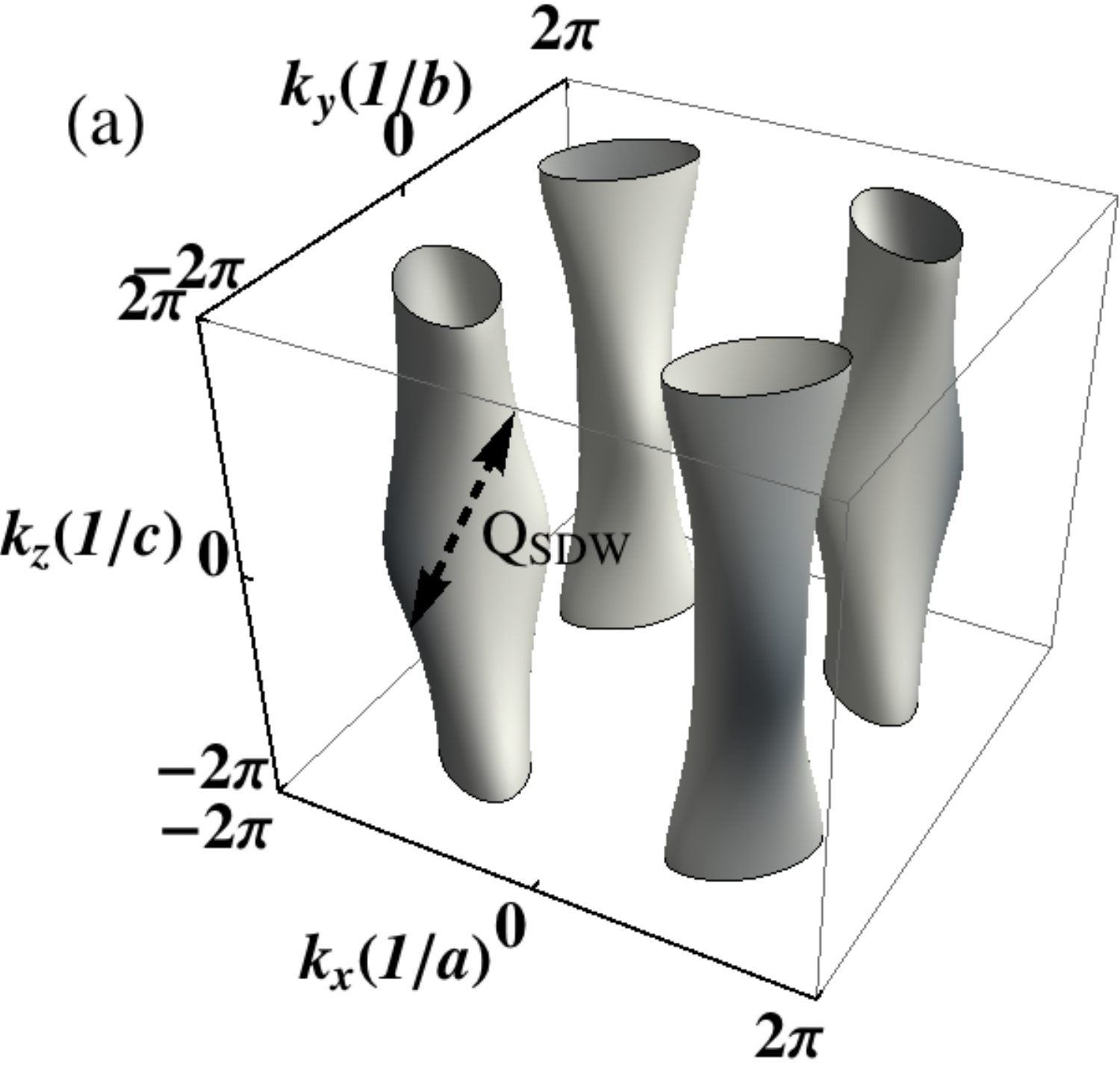}\hspace{-0.cm}
\includegraphics[width=0.52\textwidth]{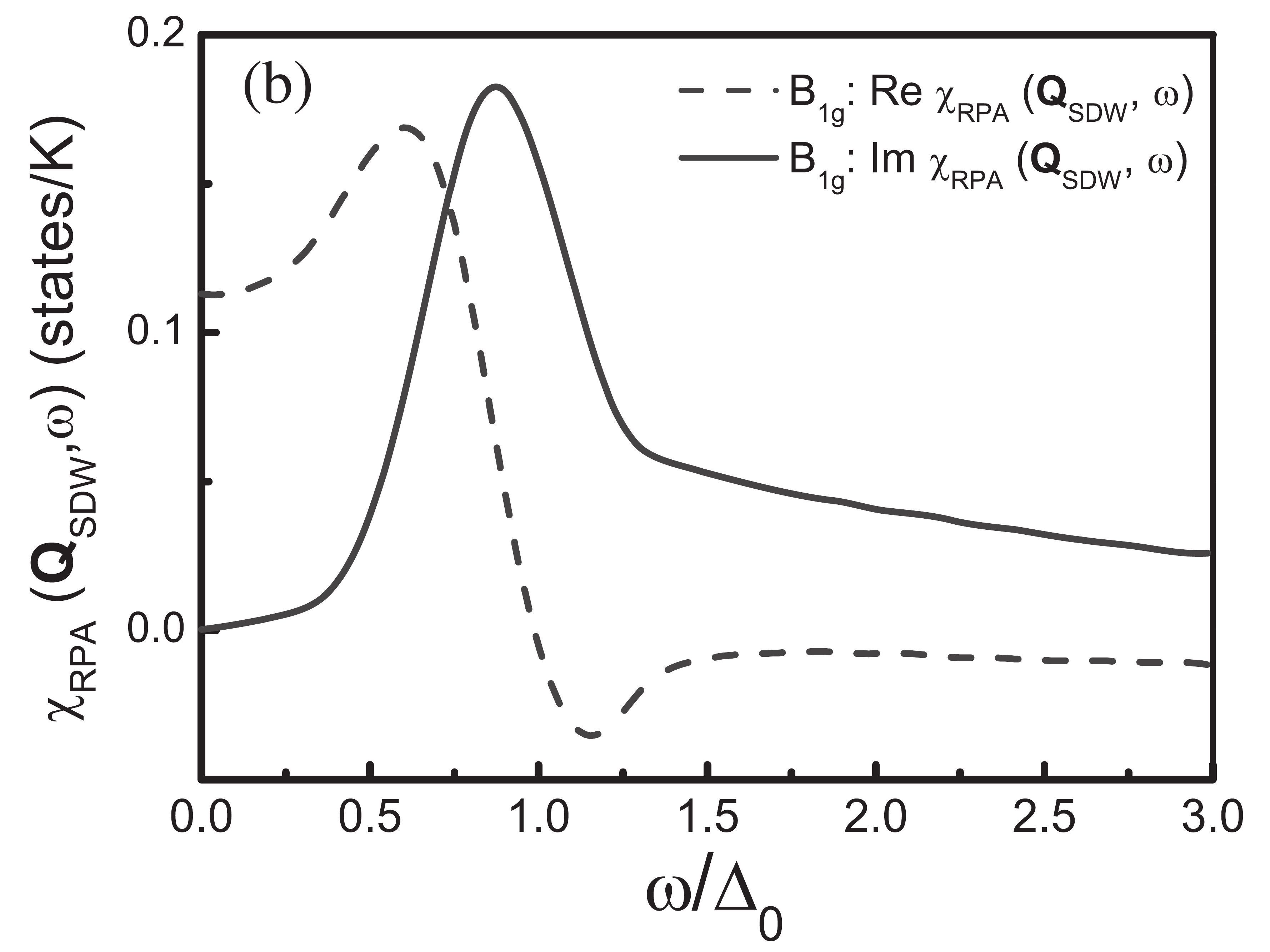}
\caption{(a) Main corrugated Fermi surface columns in \CS~from renormalized band calculations \cite{zwicknagl:92,thalmeier:05}. (b) Spin exciton spectrum and corresponding real part for the SC $d_{x^2-y^2}$ state with $\Delta_\bk=\frac{1}{2}\Delta_0(\cos k_x-\cos k_y)$ at (theoretical) IC wave vector  $\bQ=\bQ_{SDW}=(0.23,0.23,0.52)$ (Ref.~\citenum{eremin:08}).} 
\label{fig:CSres} 
\end{figure}

The observation of spin excitons in heavy fermion superconductors by INS is a challenge due to the small SC $T_c$ values and gap amplitudes $\Delta_0$ which necessarily imply a small resonance energy $\omega_r<2\Delta_0$, usually close to the experimental resolution limit. In that respect \CC~is a relatively favorable case (see Table \ref{table:resonance}) because it has the largest
T$_c$ in that class. More demanding is \CS~where the observation of a resonance finally succeeded \cite{stockert:08,stockert:11} at $\omega_r=0.2 \; \mbox{meV}$. As a major difference to the other examples in Table \ref{table:resonance} it is observed at an {\it incommensurate} wave vector equivalent to $\bQ=(0.215,0.215,0.458)$. It corresponds nicely to the nesting vector of the main heavy  Fermi surface columns of \CS~obtained in renormalized band structure calculations \cite{zwicknagl:92} shown in Fig.~\ref{fig:CSres}a. The nesting appears within each column in contrast to \CC~where different, diagonally opposite columns in  Fig.~\ref{fig:Fermi_B0} are involved. The calculated magnetic spectrum in the $B_{1g}$ $d_{x^2-y^2}$ state \cite{eremin:08} is shown in  Fig.~\ref{fig:CSres}b. 
As in \CC~it is found experimentally that the resonance is confined to the nesting vector. In the RPA expression of the spectrum this is achieved by assuming a quasiparticle interaction $J_\bq$ that decays rapidly away from \bQ~ (see Sec.\ref{sec:feedback}) such that the resonance condition Eq.~(\ref{eq:rescond}) is only fulfilled in the immediate vicinity of \bQ.  Again the resonance only appears for the $B_{1g}$ $d_{x^2-y^2}$  gap function. Even though \bQ~ is incommensurate $\Delta_{\bk+\bQ}=-\Delta_\bk$ for most of the momenta, where $\bk$ and $\bk +\bQ$ are lying now on the same FS column (Fig.~\ref{fig:CSres}a). Therefore the observation of a spin exciton resonance \cite{stockert:08} was taken as evidence for this unconventional nodal gap function in \CS \cite{eremin:08}. More recently the latter has been called into question because no specific heat oscillations in rotating field have been found and existence of a small but finite gap was concluded \cite{kittaka:14}.
 
\subsection{Satellite feedback resonance in \UP~} 
\label{subsec:UP}

\begin{figure} 
\includegraphics[width=0.49\textwidth]{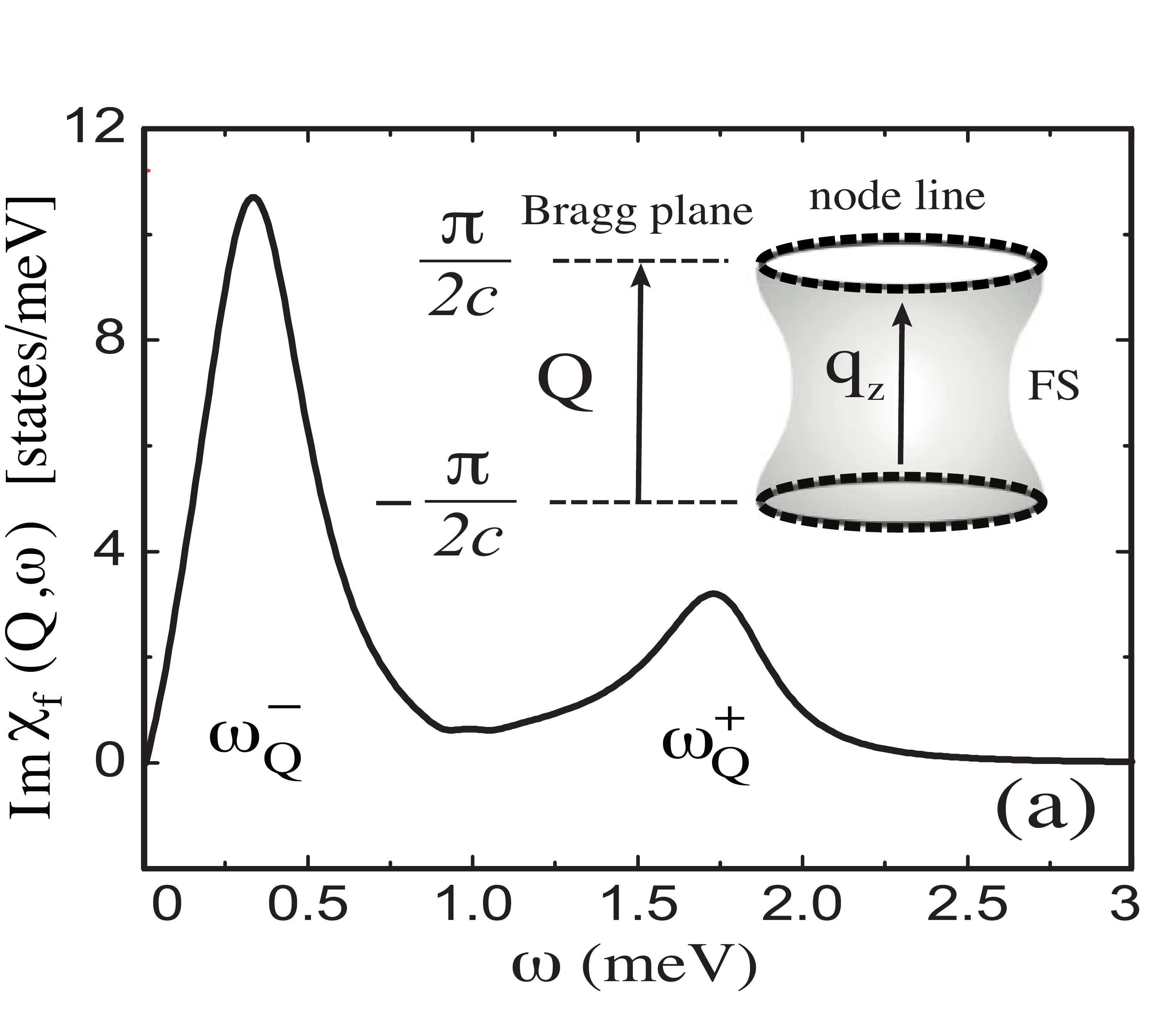}\hspace{-.cm}
\includegraphics[width=0.50\textwidth]{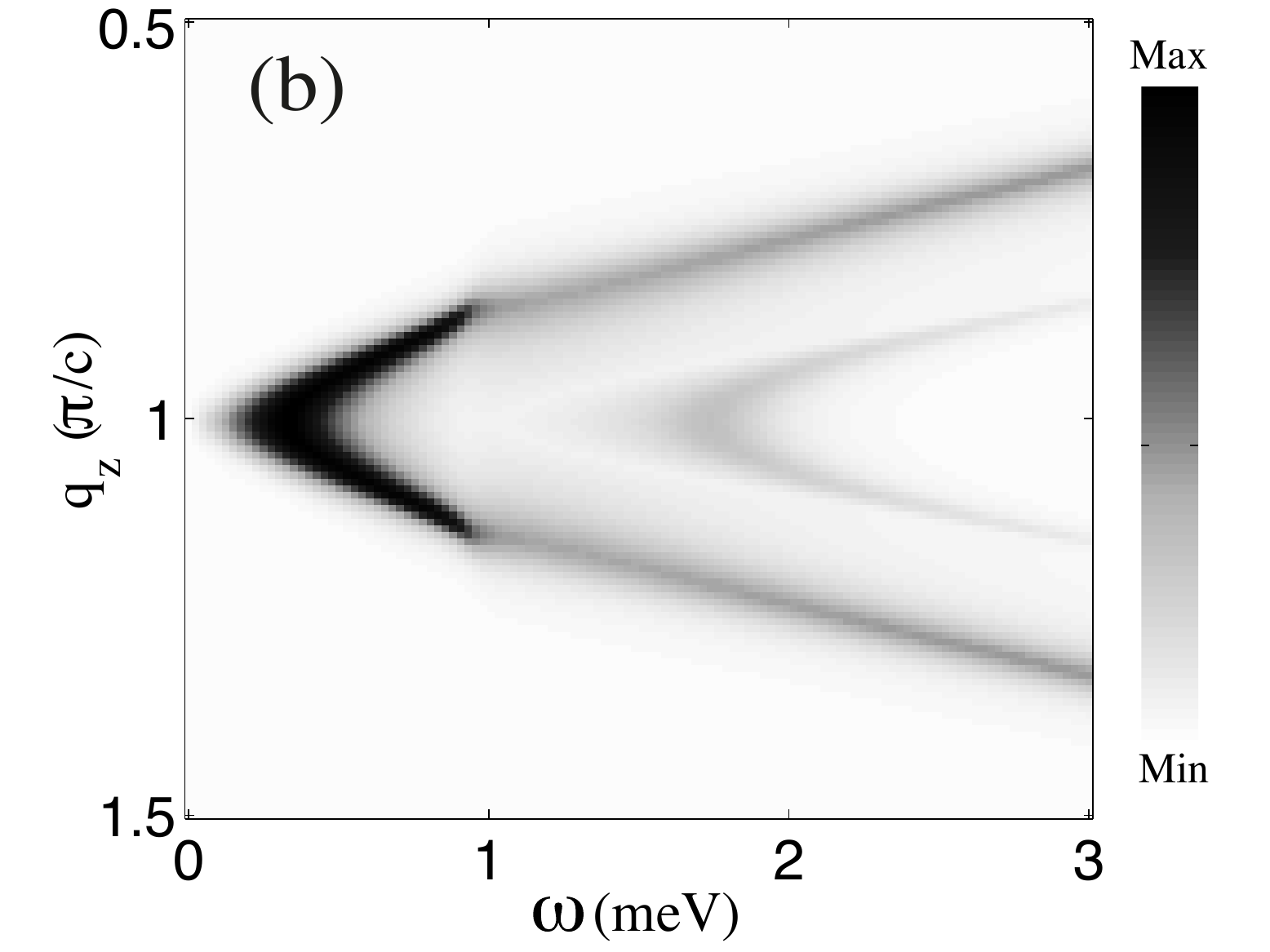}
\caption{(a) 5f-electron spectral function of \UP~in the SC state. Gap function is $\Delta_\bk=\Delta^{sc}_0\cos k_z$ with node structure on the cylindrical FS indicated in the inset. The normal state CEF singlet-singlet excitation around $\omega^+_{\bf Q}$
develops a satellite at  $\omega^-_{\bf Q}$ due to SC feedback effect.
(b) Dispersion $\omega^\pm_\bq$ of both modes for $\bq=(0,0,q_z)$ (Ref.~\citenum{chang:07}).  } 
\label{fig:UPdisp} 
\end{figure}

The third example where a feedback resonance has been observed in a heavy fermion superconductor is the hexagonal \UP~compound which  has a few special aspects: Firstly the SC phase with $T_c = 1.8$ K is embedded deeply in an AF phase (T$_N$=14.3 K) which has considerably large moments ($\mu= 0.85\mu_B$/U-site) and shows FM ordered ab planes stacked antiferromagnetically along c corresponding to $\bQ = (0,0,\frac{\pi}{c})$. Because the product of translation and time reversal is still a symmetry operation the heavy bands have a remaining effective Kramers degeneracy, enabling the coexistence with singlet pairing. The d-wave function was identified \cite{watanabe:04,thalmeier:05a,thalmeier:09} from thermal conductivity in rotating field as $\Delta_\bk=\Delta^{sc}_0\cos k_z$ which has nodal lines lying in the AF Bragg planes (inset of Fig.~\ref{fig:UPdisp}) such that the sign reversal property $\Delta_{\bk+\bQ}=-\Delta_\bk$ is fulfilled for the AF ordering vector. This raises the possibilty of  a spin exciton appearing below T$_c$ around this wave vector.
However, there is a second aspect to be considered: The heavy 5f quasiparticles in \UP~retain a partly localized character. A simplified  "dual model" of U 5f$^3$-electrons may be constructed \cite{thalmeier:02,zwicknagl:03} where two electrons are localized forming  CEF states with a small singlet-singlet splitting $\Delta\simeq 5.5\;\mbox{meV}$. The remaining one is in a quasi-2D conduction band $\epsilon_\bk$ (with little dispersion along $k_z$) that exhibits on-site exchange-coupling with strength $g$ to the virtual singlet-singlet excitations, leading to an effective mass enhancement. These disperse into a band $\omega_\bq$ of (normal state) magnetic CEF excitons between 1.5-8 \mbox{meV} by effective inter-site exchange. This model is described by
\begin{eqnarray}
\label{eq:HAMU}
H& =& \sum_{{\bf k}\sigma}\epsilon_{\bf k}c^\dagger_{{\bf k}\sigma}c_{{\bf k}\sigma} +
m_{{\bf Q}}\sum_{{\bf k}\sigma}\sigma c^\dagger_{{\bf k + Q}\sigma}c_{{\bf k}\sigma}
+\sum_{\bf q}\omega_{\bf q}\bigl(\alpha^\dagger_{\bf q}\alpha_{\bf q} + \frac{1}{2}\bigr)\non\\
&&-\frac{g}{N}\sum_{\bf k,q}c^\dagger_{{\bf k}\alpha}\sigma^z_{\alpha\beta}c_{{\bf k+q}\beta}
\lambda_{\bf q}\bigl(\alpha_{\bf q}+\alpha^\dagger_{-\bf q}\bigr)
\end{eqnarray}
where $\lambda^2_\bq=\Delta/\omega_\bq$ and $c^\dagger_\bq$,  $\alpha^\dagger_\bq$  create the conduction electron and CEF exciton, respectively. Furthermore $m_\bQ$ is the amplitude of the staggered magnetization that reconstructs the  conduction electron band in the magnetic BZ. Its cylindrical FS is shown in the inset of Fig.~\ref{fig:UPdisp}a. The dispersive CEF excitations \cite {thalmeier:02} may be described by a simplified phenomenological form as 
$\omega_{\bf q}=\omega_{ex}(1+\beta\cos q_z)$ with $\omega_{ex}\equiv\Delta$ = 5.5 meV , $\beta$ = 0.72 \cite{chang:07}. In fact INS shows that they disperse mostly along $k_z$ direction \cite{mason:97}.
In the normal state this model predicts both the mass enhancement of conduction electrons and superconducting instability due to renormalization by or exchange of bosons, respectively \cite{mchale:04}. In the superconducting state with $\Delta_\bk=\Delta^{sc}_0\cos k_z$
the coupled magnetic excitations are described by the renormalized boson propagator associated with $\alpha^\dagger_\bq$:
\begin{equation}
\label{eq:UPprop}
D({\bf q},\omega)=-\frac{2\omega_{\bf q}}{\omega^2-[\omega_{\bf q}^2-2g^2\Delta\chi_0({\bf q},\omega)]}
\end{equation}
where $\Delta\chi_0({\bf q},\omega)$ is the change in conduction electron susceptibility due to the feedback effect of superconducting gap, similar to Eq.~(\ref{eq:Lindhard}), and the AF band reconstruction. The coupled magnetic modes in the SC phase are the poles of the above propagator and approximately given by 
\begin{eqnarray}
\omega^{\pm 2}_\bQ&=&\frac{1}{2}[\omega_{\bf Q}^2+(2\Delta^{sc}_0)^2]
\pm\bigl\{\frac{1}{4}[\omega_{\bf Q}^2-(2\Delta^{sc}_0)^2]^2
+2g^2\Delta N(0)(2\Delta^{sc}_0)^2\bigr\}^\frac{1}{2}\non\\
\label{eq:UPmodes}
\end{eqnarray}
These are the approximate peak positions of the full spectrum $(-1/\pi)Im D(\bQ,\omega)$ shown in Fig.~\ref{fig:UPdisp}a.
Using the appropriate values $\Delta$ = 5.5 meV, $\omega_{\bf Q}$ = 1.54 meV , $g$ = 10 meV , 2$\Delta^{sc}_0$ = 1 meV and  $N(0)$ = 2 states/eV for conduction electron DOS we obtain the upward shifted $\omega^+_{\bf Q}$ = 1.89 meV and resonance position $\omega_r=\omega^-_{\bf Q}$ = 0.23 meV, in reasonable agreement with the peak positions of the numerical calculation in Fig.~\ref{fig:UPdisp}a. Therefore in \UP~the feedback effect leads to a satellite resonance around wave vector \bQ~ with frequency $\omega_r$ in addition to the CEF exciton $\omega^+_\bQ$ which exists already in the normal state. The latter is not present in the Ce- compounds because there are no low lying CEF excitations.
The dispersion of satellite and CEF  exciton away from the AF zone boundary (\bQ) along $q_z$ is shown in Fig.~\ref{fig:UPdisp}b. It demonstrates that both modes are dispersing in parallel and the resonance mode has the strongest intensity close to \bQ, much stronger than the intensity of the high energy CEF exciton peak. These results agree well with observations from INS \cite{sato:01,hiess:06}.

\section{Spin excitons in Kondo insulators and hidden order compounds}
\label{sec:nonSC}

We have shown that the observation of the spin exciton resonance in unconventional superconductors hinges critically on two aspects:
Firstly there should be a heavy  FS sheet with a, preferably commensurate, nesting vector \bQ~ that leads to an enhanced static spin response in $\chi_0(\bQ,0)$ already in the {\it normal} state. Secondly at this wave vector the sign change  $\Delta_{\bk+\bQ}=-\Delta_\bk$ of the SC gap function should occur which further enhances the dynamic magnetic response  $\chi_0(\bQ,\omega)$ at frequencies $\omega \approx 2\Delta_0$. This leads to a possible spin exciton bound state pole for $\omega_r<2\Delta_0$ in the collective response $Im\chi_{RPA}(\bQ,\omega)$ that shows up as sharp peaks in INS cross section. As shown in Sec.~\ref{sec:feedback} the role of the sign change is to ensure a finite coherence factor  at gap threshold which leads to a step in  $Im \chi_0(\bQ,\omega)$ around $\omega\simeq 2\Delta_0$ and a corresponding peak in $Re \chi_0(\bQ,\omega)$. This means the unconventional gap function property ensures the {\it "semiconducting" } normal state -type behavior of the magnetic response at \bQ.

Then one may question whether such spin exciton formation can be seen already directly in the normal state of Kondo lattice compounds, provided there is some kind of gap close to the Fermi energy. The gaps available are directly the hybridization gap itself and/or additional gaps introduced by hidden or magnetic order. There are indeed a few such candidates of non-superconducting Kondo compounds which clearly exhibit the spin exciton formation and they will be discussed now.

\begin{figure} 
\includegraphics[width=1.0\textwidth]{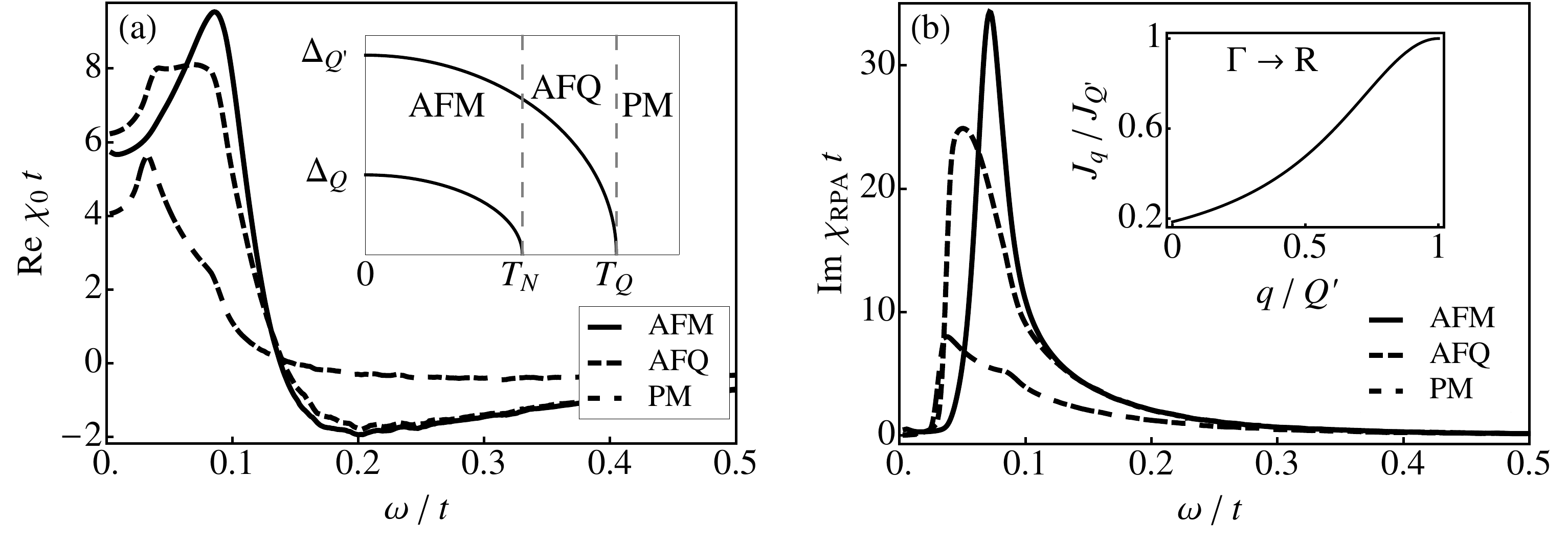}
\caption{(a) Bare susceptibility at the simple cubic (sc) R-point  $(\frac{1}{2},\frac{1}{2},\frac{1}{2})$. Due to the gaps introduced by ordering (inset) the peak value of real part strongly increases, enabling a bound state pole in $\chi_{RPA}(\omega)$. (b) Corresponding RPA spectrum with developing resonance in the ordered phase. Inset shows quasiparticle interaction $J_\bq$ along $\Gamma$R direction (Ref.~\citenum{akbari:12}).}
\label{fig:chiCeB6} 
\end{figure} 

\subsection{Spin exciton resonance in the hidden order Kondo metal \CB}

Hidden order in the bcc heavy fermion metal \CB~has a long history like that of \UR~but contrary to the latter its symmetry is well known. This may be due to the fact that 4f electrons in \CB~are more strongly localized than their 5f counterparts. In such case the possible order parameters can be classified according to the electronic multipoles of the 4f-shell \cite{shiina:97,kuramoto:09}.
 The  $Ce^{3+}$ $J=5/2$ multiplet of 4f electrons is split by the cubic CEF into fourfold $\Gamma_8$ ground state and a $\Gamma_7$ excited state. The latter is neglected due to its large splitting energy of $46\;\mbox{meV}$. The ground state carries 15 multipoles, namely 3 dipoles (rank 1), 5 quadrupoles (rank 2) and 7 octupoles (rank 3) where the even($+$)/odd($-$) rank multipoles break/conserve time reversal symmetry. It was proposed theoretically \cite{shiina:97} that the primary order parameter is a $\Gamma^+_5$ quadrupole (T$_Q$=3.2 K) and the field induced order parameter a $\Gamma^-_2$ octupole both of anti-ferro type with wave vector $\bQ'=(\frac{1}{2},\frac{1}{2},\frac{1}{2})$. They have been identified by neutron diffraction \cite{erkelens:87} and resonant x-ray scattering (RXS) \cite{matsumura:09}. At a lower temperature T$_N$=2.3 K a further transition to a dipolar AF phase with a different  $\bQ=(\frac{1}{4},\frac{1}{4},0)$ is observed.
 
The localized 4f multipole model explained the thermodynamics and H-T phase diagram \cite{shiina:97}, NMR characteristics \cite{shiina:98}, RXS results \cite{nagao:01} and partly high-field INS results \cite{shiina:03,thalmeier:03}. It is, however, surprising that a localized approach should cover all aspects of \CB~since this compound is a model Kondo lattice systems with a Kondo temperature T$_K=4.5$ K \cite{loewenhaupt:85}  corresponding to the width of the heavy quasiparticle band with a mass enhancement $m^*/m_e\simeq 20$ ($\gamma =250 mJ/molK^2$). Therefore the quasiparticle band width T$_K$ is of the same size as the ordering temperature T$_Q$ which might suggest that the itinerant nature of 4f electrons should manifest itself also in the HO phase.  
\begin{figure} 
\hspace{0.70cm}
\includegraphics[width=0.85\textwidth]{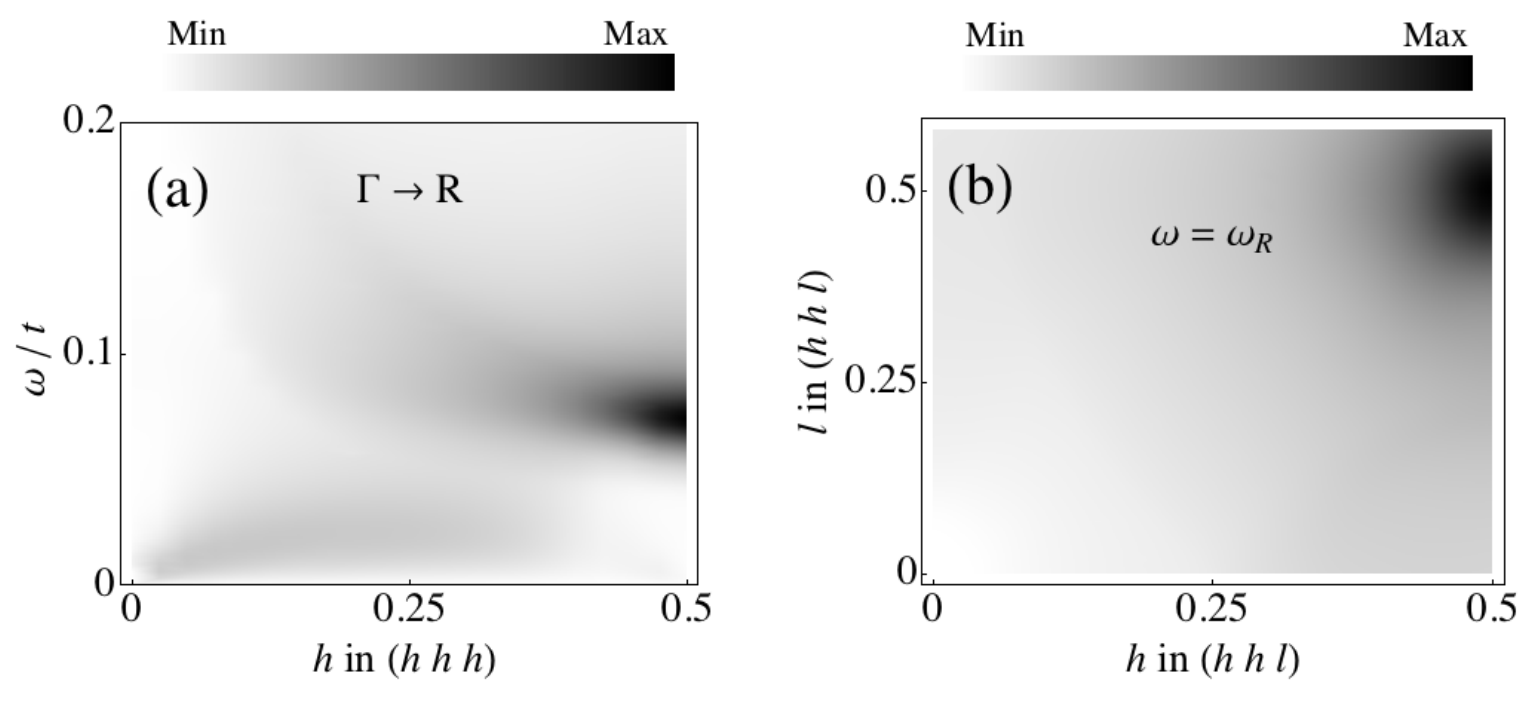}
\caption{(a) Contour plot of $ Im\chi_{RPA}({\bf q},\omega)$ along $\Gamma$R direction. Localized resonance peak appears at R for $\omega=\omega_r$ and an associated spin gap develops below. (b) Constant energy ($\omega=\omega_r$) scan in the full (hhl) scattering plane (Ref.~\citenum{akbari:12}).} 
\label{fig:contCeB6}
\end{figure} 
In fact subsequent INS experiments \cite{friemel:12,jang:14} have seriously raised this issue. They have shown that \CB~exhibits a magnetic low  energy mode in zero field that has all the basic features of an itinerant spin exciton resonance in the unconventional superconductors:
i) it has a sharp energy of $\omega_r=0.5\;\mbox{meV}$, ii) it is confined to the AFQ momentum \bQ' ( simple cubic R-point).
iii) The temperature dependence of $\omega_r$ and its intensity increase in an order-parameter like fashion with decreasing temperature and simultaneously  the intensity for $\omega <\omega_r$ is depleted, i.e. a spin gap forms. Since there is no superconducting feedback effect one must conjecture that the hybridization gap and the additional gaps introduced by the orderings lead to the necessary singular behavior of the bare magnetic susceptibility to allow for a bound state. This may be described by the mean field hybridization model
of Eq.~(\ref{eq:HAMF}) supplemented by the effect of the molecular fields due to AFQ and AFM order which lead to the additional gapping of the mean field quasiparticle spectrum\cite{akbari:12}. These terms are given by
\bea
{\cal H}_{AFQ}=\sum\limits_{{\bf k}\sigma} 
\Delta_{{\bf Q}^\prime}(f_{{\bf k},+\sigma }^{\dagger}f_{{\bf k}+{\bf Q}^\prime-\sigma }+f_{{\bf k}, -\sigma}^{\dagger}f_{{\bf k}+{\bf Q}^\prime,+\sigma}).
\label{eq:AFQ}
\eea
and
\bea
{\cal H}_{AFM}=\sum\limits_{{\bf k} \tau} \Delta_{\bf Q}
(f_{{\bf k} \tau\uparrow }^{\dagger}f_{{\bf k}+{\bf Q} \tau\downarrow }+f_{{\bf k} \tau\downarrow }^{\dagger}f_{{\bf k}+{\bf Q} \tau\uparrow })
\label{eq:HAFQ-AFM}
\eea
Here $\tau=\pm$  denotes the pseudo-orbital and $\sigma=\pm$ the (Kramers-) pseudo spin degree of freedom of the $\Gamma_8$ quartet states.  To find the magnetic excitations we need the  dipolar susceptibility
 $
 \chi_{0}^{ll^\prime}({\bf q},t)=-
 \theta(t)
 \bra
 T j_{{\bf q}}^{l}(t)j_{-{\bf q}}^{l^\prime}(0)
 \ket,
 $
 where
$
  j_{{\bf q}}^{l}=\sum\limits_{{\bf k}mm^\prime} 
  f_{{\bf k}+{\bf q}m} ^{\dagger}
  {\hat M}^{l}_{mm^\prime} 
  f_{{\bf k}m^\prime}
$ 
are  the physical magnetic dipole operators ($l,l^\prime=x,y,z$)
with $ {\hat M}^{z}= (7/6)\hat{\tau}_0 \otimes \hat{\sigma}_z$.
Due to cubic symmetry we can restrict to $ \chi_{0}^{zz}({\bf q},\omega)$ given by ($i\nu\rightarrow \omega+i 0^+$)
 \bea
 \chi_{{\bf q}}^{zz}(\omega)
\propto
 \sum\limits_{\alpha \alpha^\prime{\bf k}m_1m_2} 
   ( \hat{ \rho}_{{\bf  k},{\bf q} } ^{\alpha^\prime \alpha})^2
   \int d\omega^\prime
   {\hat G}^{0}_{ss}(i\nu+\omega^\prime)      {\hat G}^{0 }_{s^\prime s^\prime }(\omega^\prime)
 \eea
where we defined $s= (\alpha,{{\bf  k}+{\bf q}}, m_1 )$ and  $s^\prime=(\alpha^\prime,{{\bf  k}}, m_2)$.
It contains the effect of the new quasiparticle energies in the Green's functions  ${\hat G}^{0 }_{s s}$ and matrix elements  $\hat{ \rho}_{{\bf  k},{\bf q} } ^{\alpha^\prime \alpha}$  reconstructed by the molecular fields of Eqs.~(\ref{eq:AFQ},\ref{eq:HAFQ-AFM}), similar to the coherence factors in the superconducting case.
The effect of this reconstruction due to AFQ/AFM gap openings is seen in Fig.~\ref{fig:chiCeB6}a.
The magnetic response is pushed to higher frequencies and the  real part is considerably enhanced in the ordered phase.
As a consequence the RPA susceptibility 
\bea
 \chi_{RPA}({\bf q},\omega)=
 [1- J_{{\bf q}}  \chi_{0}^{zz}({\bf q},\omega)]^{-1} \chi_{0}^{zz}({\bf q},\omega),
 \label{eq:RPA}
\eea
develops a bound state pole due to the effect of the ordering. This is seen from 
 Fig.~\ref{fig:chiCeB6}b where below $T_Q$ and in particular $T_N$ a sharp resonance evolves around the AFQ ordering vector $\bq\approx \bQ'$ at an energy  $\omega_r/\Delta_c =0.64$  for $T\rightarrow 0$.  Here $\Delta_c$ is the indirect hybridization charge gap
(Table \ref{table:resonance}) observed in point contact spectroscopy \cite{paulus:85} and illustrated by the DOS in  Fig.~\ref{fig:Quasiband}b.
The momentum dependence of the spectrum along BZ diagonal $\Gamma$R-line is shown in 
Fig.~\ref{fig:contCeB6}a. It demonstrates the confined resonance excitation at $\omega_r$ 
and the existence of the spin gap $(\omega\ll\omega_r)$, both at the R-point. Away from the R-point
the low energy spin fluctuations of the metallic state are present. The complementary Fig.~\ref{fig:contCeB6}b
shows the magnetic intensity  in the $(hhl)$ scattering plane precisely at $\omega=\omega_r$. It piles  up at
the resonance location R and decays rapidly away from it. The similar plot for $\omega\ll\omega_r$ would 
show the {\it "negative"} of that figure with very low intensity at the spin gap region around R.

\subsection{Dispersive  spin exciton mode doublet in the hybridization-gap semiconductor \YB}
\label{subsec:YB}

In \CB~the resonance is tied to the appearance of the hidden and AF order that enhance
$\chi_{0}^{zz}({\bf q},\omega)$ which enables the existence of a pole in Eq.~(\ref{eq:RPA}). One might, however,
suspect that this is not apriori necessary and under favorable conditions the resonance appears already without
the support of additional gapping by the effect of order. This case is realized in cubic \YB~\cite{akbari:09}.
The compound is a true Kondo semiconductor with equal spin and charge gap of $\Delta_c \sim 15\;\mbox{meV}$
\cite{,nemkovski:07} and without any ordering. The 4f-hole in $Yb^{3+}$ has a lowest $J=7/2$ multiplet which is split by the CEF into a
$\Gamma_8^{(1)}$ ground state and two close by doublets which we simplify as another pseudo-quartet $\Gamma_8^{(2)}$.
\begin{figure} 
\includegraphics[width=0.54\textwidth]{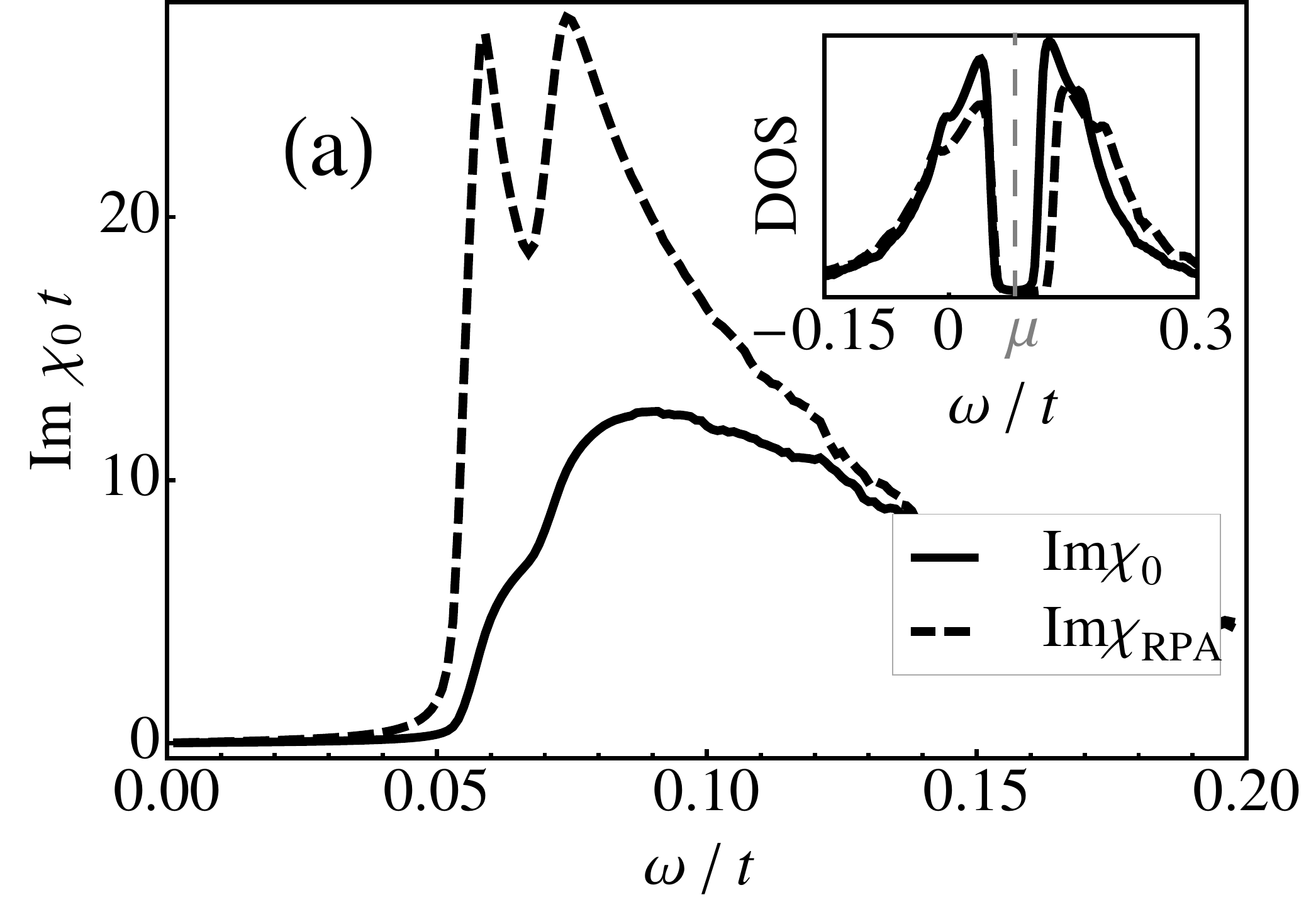}\hfill
\includegraphics[width=0.44\textwidth]{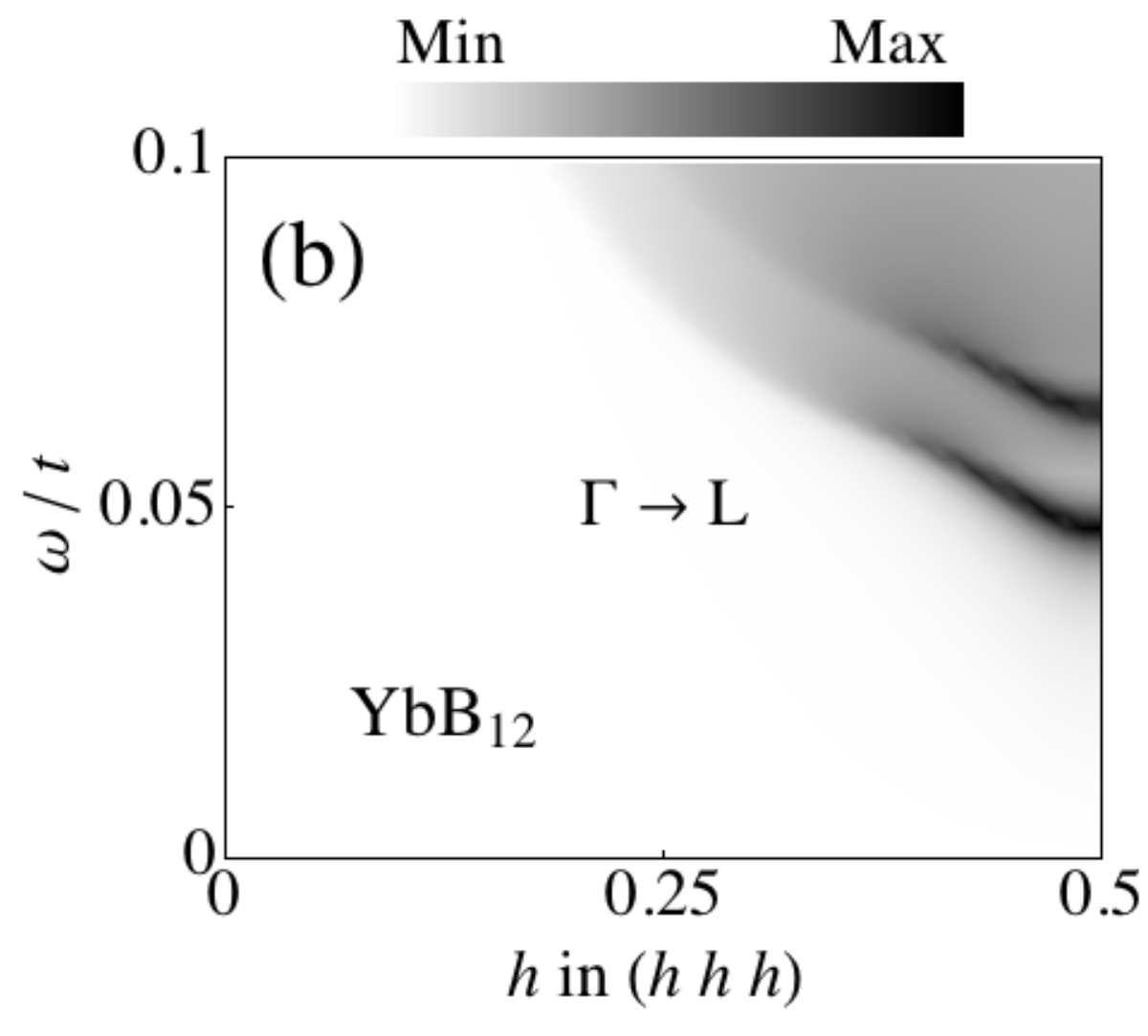}
\caption{(a) Inset: different hybridization leads to different size of quasiparticle hybridization gaps.  Then the resonance conditions result  in a split $\omega_r$ position (main panel). (b) Dispersion of split resonance along $\Gamma$L direction where L is the bcc  $(\frac{1}{2},\frac{1}{2},\frac{1}{2})$ point (Ref.~\citenum{akbari:09}).
 } 
\label{fig:contYbB12} 
\end{figure} 
Then the model in Eq.~(\ref{eq:HAMF}) has to be slightly generalized  according to $\epsilon_f \rightarrow \epsilon_f+\Delta_\Gamma$
and $V\rightarrow V_\Gamma$ to include the CEF splitting $\Delta_2-\Delta_1$  of the two quartets and in particular their different average hybridization $V_\Gamma=\frac{1}{2}\bigl(\sum_m|V_{\Gamma m}|^2\bigl)^\frac{1}{2}$. Then we obtain two sets of quasiparticle bands with different size of the hybridization gap as seen in the inset of Fig.~\ref{fig:contYbB12}. The corresponding bare susceptibility is now given by
\bea
\chi_0^{\Gamma}({\bf  q},\omega)
&=&\sum_{{\bf  k},\pm}u^{\Gamma}_{\pm \bk+\bq} u^{\Gamma}_{\mp \bk}
\left[\frac{f(E^\pm_{\Gamma}({\bf  k}+{\bf  q}))-f(E^\mp_{\Gamma}({\bf  k}))}
{E^\mp_{\Gamma}({\bf  k})-E^\pm_{\Gamma}({\bf  k}+{\bf  q})-\omega}\right],
\eea
which then implies that the collective RPA susceptibility has contributions from the two sets of bands
according to
\bea
\chi_{RPA}({\bf  q},\omega)=\sum_{\Gamma}[1-{\it J}_{\Gamma}({\bf  q})
\chi_0^{\Gamma\Gamma}({\bf  q},\omega)]^{-1}\chi_0^{\Gamma\Gamma}({\bf  q},\omega).
\eea
where the effective interaction ${\it J}_{\Gamma}({\bf  q})$ will also depend on the quartet state.
This means that one obtains two split collective modes with different energies  if the resonance condition is
fulfilled for each $\Gamma$. This is shown in Fig.~\ref{fig:contYbB12}a where two resonance 
peaks appear at the $\bQ =(\frac{1}{2},\frac{1}{2},\frac{1}{2})$ bct L-point  right on top of the single particle hybridization gap. This wave vector corresponds to the low energy indirect interband ($\pm$) excitations across the hybridization gap (see illustration in Fig.\ref{fig:Quasiband}). When the momentum decreases away from the L-point the interband excitation energies increase
from the indirect band gap $\sim T_K$ at \bQ~ to the direct band gap $\sim 2\tilde{V}_\Gamma \gg T_K$ for $\bq\rightarrow 0$. This leads to a decrease of $Re \chi_0^{\Gamma\Gamma}({\bf  q},\omega)$ for fixed $\omega$ and therefore the resonance condition becomes harder to fulfill for both modes.
Finally at about one third into the BZ the intensity of the spin excitons vanishes as seen in  Fig.~\ref{fig:contYbB12}b. The upward dispersion of the split modes is again due to the behavior of $Re \chi_0^{\Gamma\Gamma}({\bf  q},\omega)$ whose maximum in $\bq,\omega$- plane moves to larger energies with decreasing $|\bq|$, this also results in larger energies of the two resonances (Fig.~\ref{fig:contYbB12}b).
The model parameters have been adapted to reproduce the hybridization gap, the observed resonance energies and the dispersive features. It is interesting to speculate what would happen if they could be tuned physically. A decrease of the former (or an increase in ${\it J}_{\Gamma}({\bf  q}$))  might in principle lead to a soft spin exciton mode at the L-point producing an antiferromagnetic Kondo insulator.

\subsection{Commensurate spin exciton resonance in the hidden order phase of \UR}
\label{subsec:UR}

The most well studied f electron compound with HO below $T_{HO}=17.8\;\mbox{K}$ is tetragonal \UR. 
There is no space here to recount the long history of this subject \cite{mydosh:11}.
We restrict ourselves to describe some striking relations between the appearance of HO and low energy INS results in this compound \cite{villaume:08,bourdarot:10}. It is generally accepted that the itinerant nature of U 5f states involved in HO cannot be neglected \cite{oppeneer:10}.
An ab-initio treatment of itinerant 5f bands and their possible multipolar HO instabilities was given in Ref.~\citenum{ikeda:12}, see also Ref.~\citenum{thalmeier:14}. A simplified toy model \cite{rau:12} to describe 5f bands is useful to illustrate the meaning of itinerant HO. 
The states on the Fermi surface belong to the two $\Gamma_7^{(1),(2)}$ doublets of $j=5/2$ multiplet. These  jj-coupled single particle states are described by the basis
\bea
\left(
\begin{array}{c}
f_{1\pm}\\
f_{2\pm}
\end{array}
\right)
=
\left(
 \begin{array}{cc}
\cos\theta& \sin\theta \\
 -\sin\theta& \cos\theta
\end{array}
\right)
\left(
\begin{array}{c}
f_{\pm\frac{5}{2}}\\
f_{\mp\frac{3}{2}}
\end{array}
\right)
\label{eq:CEFtrans}
\eea
with CEF splitting $\Delta_f$ and mixing angle $\theta$ determined by the tetragonal CEF parameters. Within this basis the kinetic energy of heavy 5f quasiparticles is  \cite{rau:12}:
\bea
&&H_0=\sum_{\bk\si}\bigl(A_{1\bk}f^\dg_{1\si\bk}f_{1\si\bk}+A_{2\bk}f^\dg_{2\si\bk}f_{2\si\bk}\bigr)\non\\
&&+\sum_\bk\bigl[D_{\bk}\bigl(f^\dg_{1+\bk}f_{2-\bk}-f^\dg_{2+\bk}f_{1-\bk}\bigr)
+D^*_{\bk}\bigl(f^\dg_{2-\bk}f_{1+\bk}-f^\dg_{1-\bk}f_{2+\bk}\bigl)\bigr]
\label{eq:bands}
\eea
%
\begin{figure} 
\includegraphics[width=1\textwidth]{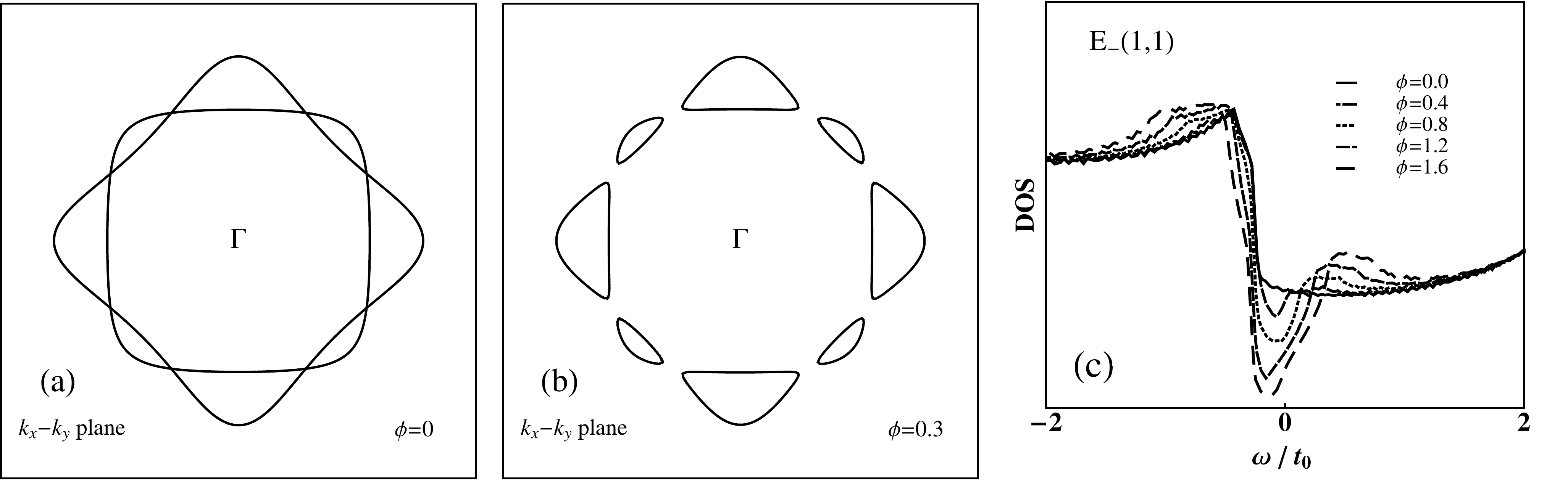}
\caption{(a) $\Gamma$ centered electron (diagonal oriented) and projected $Z (0,0,2\pi/c)$ centered hole (axis oriented) Fermi surface sheets in the $k_x-k_y$ plane for the para phase. Nesting with $\bQ= (0,0,2\pi/c)$ appears in the crossing region. (b) Gapping of the Fermi surface in the nesting region within the HO phase ($\phi=0.3$) (using model parameters of Ref.~\citenum{rau:12}). (c) Corresponding DOS showing the evolution of HO gap around $\omega\approx 0$ which leads to the low energy spin resonance inside the HO charge gap $\Delta_c=0.61t_0$  ($t_0=6.7$ meV). 
 } 
\label{fig:URFS} 
\end{figure}
%
where $A_{1,2\bk}, D_\bk$ are model functions parametrized with effective f-orbital energies and  f-hopping matrix elements. This leads to four effective 5f-electron bands in the simple tetragonal BZ (downfolded by HO vector \bQ, using $D_{\bk+\bQ}=-D_\bk$):
\bea
\varepsilon^\pm_1(\bk)&=&
\frac{1}{2}(A_{1\bk}+A_{2\bk}) \pm\bigl[\frac{1}{4}(A_{1\bk}-A_{2\bk})^2+|D_\bk|^2\bigr]^\frac{1}{2}\non\\
\varepsilon^\pm_2(\bk)&=&
\frac{1}{2}(A_{1\bk +\bQ}+A_{2\bk +\bQ}) \pm\bigl[\frac{1}{4}(A_{1\bk +\bQ}-A_{2\bk +\bQ})^2+|D_\bk|^2\bigr]^\frac{1}{2}\
\label{eq:UPbands}
\eea
The resulting Fermi surface cut with $k_z=0$, consisting of electron and hole pockets around bct $\Gamma, Z$ points and projected into the simple tetragonal BZ,  is shown in Fig.~\ref{fig:URFS}a. Due to the electron-hole nesting property for wave vector \bQ~  an instability may occur. According to Ref.~\citenum{ikeda:12} the dominating instability corresponds to a multipolar HO. It is proposed \cite{shibauchi:14}  that HO breaks i) translational symmetry (HO is of antiferro type with $\bQ=(0,0,1))$ ii) rotational $C_4$ in-plane symmetry and iii) presumably time reversal invariance. The most likely candidate is then a rank-5 $E_-$ "dotriacontapole" which leads to a molecular field term in the Hamiltonian given by ($\kappa=1/(2\sqrt{2})$)
\bea
E_{-}: H_\phi&=&-\kappa\phi_\bQ\cdot\sum_\bk(f^\dg_{1\bk}\sigma f_{2\bk+\bQ}+f^\dg_{2\bk}\sigma f_{1\bk+\bQ}) +H.c.
\label{eq:HOP}
\eea
In the original $f_M$ $(|M|\leq\frac{5}{2})$ CEF states used in Eq.~(\ref{eq:CEFtrans}) this corresponds to electron-hole pairing where the pairs have a maximum angular momentum difference $\Delta M=\pm 5$ equal to the rank.
Here we introduce the HO vector $\phi_\bQ=(\phi_x,\phi_y)$ which expresses the twofold degeneracy of $E_{-}$ representation, furthermore $\sigma =(\sigma_x,\sigma_y)$. Therefore right at T$_{HO}$ the HO phase has continuous $U(1)$ symmetry which is lifted by higher order terms in the free energy below T$_{HO}$, selecting a direction for $\phi_\bQ$. From torque experiments \cite{okazaki:11,thalmeier:14,shibauchi:14} it is concluded that  a phase with equal components $\phi_x=\phi_y$ called $E_-(1,1)$ phase is realized. It reconstructs the Fermi surface by gapping out the states connected by the nesting vector (the crossing region in Fig.~\ref{fig:URFS}a,b). This leads to a strong reduction in charge carrier DOS at the Fermi level with increasing $\phi_\bQ$ (Fig.~\ref{fig:URFS}c)
 and the typical singular bare magnetic response for frequencies around the HO gap $\Delta_c$ similar to \CB.
Then it is suggestive that a collective spin exciton resonance  will appear within the gap at the nesting or HO vector \bQ. This has indeed been found in polarized INS experiments \cite{villaume:08,bourdarot:10}. It was shown that: i) The resonance only appears in the longitudinal ($\parallel$ z-axis) response at an energy $\omega_r = 1.86 \;\mbox{meV}$ corresponding to $\omega_r/\Delta_c=0.45$ where $\Delta_c=4.1 \; \mbox{meV}$ is the HO charge gap obtained from STM results \cite{aynajian:10}. ii) The integrated intensity of the resonance shows a clear order parameter (BCS)-type temperature dependence below $T_{HO}$. iii) The resonance vanishes above the critical pressure when HO is destroyed. This proves clearly that the commensurate spin exciton resonance at \bQ~ is closely linked to the appearance of FS gapping caused by hidden order. 

\section{Feedback effect on magnetic excitations in Ce-based Fe- pnictide superconductors}
\label{sec:Fe}

The Fe-pnictides are not strongly correlated systems, nevertheless we discuss here one example where an indirect superconducting feedback effect on Ce 4f excitations was found in \CF~($x=0.16$ and T$_c=41$K) \cite{chi:08}.
In  Fe-pnictide superconductors the unconventional $s_\pm$ gap function changes sign between the $\Gamma (0,0)$ centered hole and M ($\pi,\pi$) -centered electron pockets. The sign change was proven directly by the phase sensitive QPI method in magnetic fields \cite{hanaguri:10}. Already before it was suggested that the $s_\pm$-state should lead to the appearance of a spin exciton resonance below T$_c$ at the AF nesting wave vector \bQ=($\pi,\pi$) (folded BZ) \cite{korshunov:08}. This has indeed been found  in many Fe-pnictide systems \cite{lumsden:09,inosov:10}, giving strong support for the $s_\pm$ nature of the gap function. The spin exciton in this case is to be viewed as a collective Fe-3d electron magnetic mode in the 2D superconducting FeAs planes.\\

In \CF~,however, there are {\it two} magnetic subsystems present which are a large  distance 
apart. These are the tetrahedrally coordinated FeAs-planes with 3d electrons separated by spacer layers of CeO pyramides with 4f electrons. The 3d electrons are in itinerant conduction states that interact via Coulomb repulsion, leading to spin fluctuations, superconductivity and the feedback spin exciton. The well localized 4f electrons in Ce layers have negligible direct intersite interaction among themselves, but due to the extended 3d wave functions the 4f moments will be weakly coupled to the 3d layers via exchange interaction.
Therefore the Ce 4f moments  may act as probes to the magnetic dynamics of the 3d subsystem. If the latter develops a collective spin exciton below T$_c$ this should have an effect on the 4f magnetic excitations due to the Fe-Ce interlayer exchange.  Furthermore the tetragonal CEF splits the $J=5/2$ 4f states into three Kramers doublets. We restrict to the ground state $|0\rangle$ ($\varepsilon_0 = 0$) and first excited state $|1\rangle$ doublet at $\varepsilon_1\equiv \Delta_f=18.7$ meV. Then the interlayer exchange coupling will influence the 4f-CEF transition with energy $\Delta_f$, affecting both position and line width. One would expect that the latter strongly decreases in the SC state due to the opening of the gap and reduction of Landau damping effects on localized $\Delta_f$ excitations. Surprisingly the opposite was found \cite{chi:08}: The 4f CEF line width shows a considerable increase, indicating an anomalous spin dynamics in the 3d-electron system to which it is coupled. This remarkable observation was explained by a two component model \cite{akbari:09a} defined by
\bea 
H =  \sum_{i\gamma}\varepsilon_{\gamma}\left| i\gamma \rangle
\langle i \gamma \right|+\sum\limits_{{\bf k} ,\sigma
} {\varepsilon_{{\bf k}}} d_{{\bf k} \sigma }^\dag d_{{\bf k}
\sigma}+U\sum_{i,m}n_{di\uparrow}n_{di\downarrow}
-I_{0}\sum_{i}{\bf s}_{i}{\bf J}_i
 \label{eq:Hpnic}
  \eea
The first term describes isolated CEF split ($\gamma =0,1$) 4f states in Ce layers, the second and third term describe the correlated 3d electrons in FeAs layers and the last term presents the interlayer exchange between localized 4f moments with total angular momentum ${\bf J}_i$ and itinerant 3d spins ${\bf s}_i$.  Without the last term the separate magnetic response in 4f, 3d layers is given, respectively, by
%
\begin{figure} 
\hspace{0.1cm}
\includegraphics[width=0.98\textwidth]{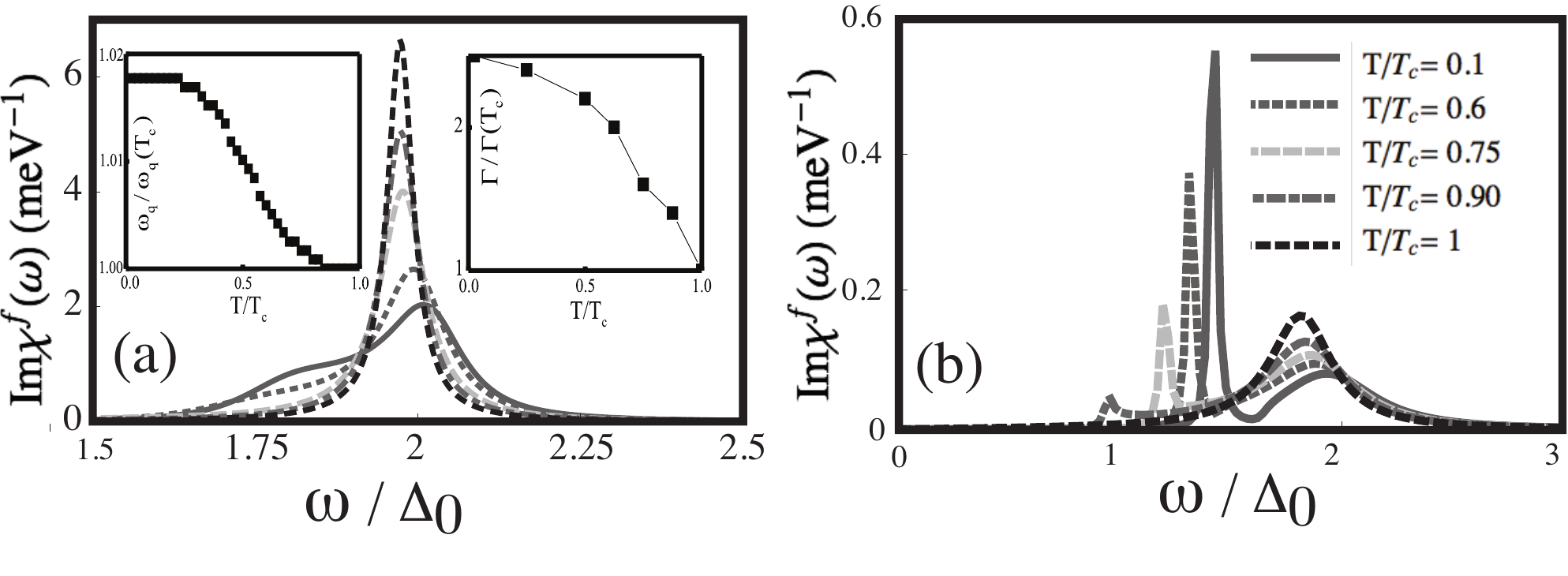}
\caption{
Evolution of spectral function of f-electrons with CEF splitting $\De_f\simeq 2\De_0$ as function of f-d coupling strength.  Curves correspond to $T/T_c=1.0, 0.5, 0.75, 0.9, 0.1$ for decreasing peak height. (a) for small coupling  $I_0\simeq 0.72\De_0$ an upward shift of the peak position $\omega_q(T)$ (left inset) and increase in line width $\Gamma(T)$ (right inset) are observed. 
(b) For larger f-d coupling  $I_0\simeq \De_0$ a double peak structure with a split-off satellite at lower energy develops (Ref.~\citenum{akbari:09a}).} 
\label{fig:Fepnic}   
\end{figure}
%
%
\begin{eqnarray}
u_{\alpha}(\omega)&=&|m_{\alpha}|^2\frac{2\Delta_{f}\tanh(\Delta_{f}/2T)}
{(\Delta_{f}^2-\omega^2)};\;\;
\chi^{(d)}_{RPA} ({\bf q}, \omega)=  \frac{\chi^{(d)}_0 ({\bf q},
\omega)}{1-U \chi^{(d)}_0 ({\bf q}, \omega)}
\label{eq:RPAfd}
\end{eqnarray}
Here $u_{\alpha}(\omega)$ is susceptibility of an isolated 4f CEF doublet-doublet system with splitting $\Delta_f$ and dipolar matrix element $m_\alpha^2=|\langle 0| J_\alpha | 1 \rangle |^2$ ($\alpha=\perp,\parallel$) and $\chi^{(d)}_{RPA}$ is the collective susceptibility of the 3d subsystem in the SC state with $s_\pm$ gap function $\De_\bk=(\De_0/2)(\cos k_x +\cos k_y)$. It will develop the spin exciton pole at $\omega_r <2\Delta_0$ at $\Gamma \mbox{M}$ nesting vector $\bQ=(\pi,\pi)$  below T$_c$ because of $\De_{\bk+\bQ}=-\De_\bk$. For the coupled system including the last term in Eq.~(\ref{eq:Hpnic}) the f-electron response is then described by
\begin{eqnarray}
\chi^f(\bq,\omega)&=&\frac{u(\omega)}{1-I_0^2u(\omega)\chi^d_{RPA}(\bq,\omega)}
\label{eq:RPApnic}
\end{eqnarray}
Its spectral function may be characterized by the position and damping of the pole in this expression which are obtained as
\begin{eqnarray}
\omega_{\bf q}^2&=&\Delta_{f}^2-2\Delta_{f}
(I_0m_\perp)^2\left[\chi_{RPA}^{(d)}({\bf q},\omega_{\bf q})\right]^{\prime}\non\\
\Gamma_\bq&=&2(I_0m_\perp)^2\Delta_f\bigl[\chi^{(d)}_{RPA}(\bq,\omega_\bq)\bigr]^{''}
\label{eq:modepnic}
\end{eqnarray}
The selfconsistent solution of the first equation describes the shifted CEF excitation $\omega_\bq$ of Ce around $\Delta_f$ and the second its line width $\Gamma_\bq$, both
caused by the exchange coupling $\sim I_0$ to the itinerant 3d electrons. These effects are directly proportional to the 3d conduction electron magnetic response in the FeAs layer. If the latter develops a spin exciton resonance below T$_c$ it will be probed by level shift and line width of the CEF excitation. This has indeed been observed experimentally \cite{chi:08}. The theoretical results are presented in  Fig.~\ref{fig:Fepnic}a showing the evolution of the 4f spectral function around $\Delta_f$ as function of temperature. The insets depict the corresponding position and line width dependence of the renormalized CEF peak on temperature. Obviously the line width {\it increases} below $T_c$, opposite to the behavior in conventional superconductors, because of the coupling to the resonant feedback effect in the superconducting 3d subsystem. In Fig.\ref{fig:Fepnic}a we use a subcritical $I_0 < \Delta_0$ which leads to shift and broadening of the CEF excitation in qualitative agreement with experiment.
However, it is instructive to consider also the critical case $I_0 \simeq \Delta_0$. Then the first of Eq.~(\ref{eq:modepnic}) will have two solutions, the shifted $\Delta_f$ level and a new sharp satellite at lower energy. The resulting 4f- spectral function is shown in Fig.~\ref{fig:Fepnic}b as function of temperature. Approximately the two peak positions are given by
\bea
\label{eq:splitpole} 
\omega_q^{\pm} \simeq  \frac{1}{2}(\Delta_{f}+ \omega_r)
\pm \frac{1}{2}\sqrt{\left(\Delta_{f}-
\omega_r\right)^2+4I_0^2|m_{\perp}|^2z_r}
\eea
where $\omega_r, z_r$ are position and residue of the resonance peak in the $3d$ spin
susceptibility. This case is remarkably similar to the previously discussed 5f electron spectral shape in \UP.
It also develops a shifted CEF and additional satellite peak in the SC state described by Eq.~(\ref{eq:UPmodes}) and shown in 
Fig.~\ref{fig:UPdisp}. However, the physical situation is qualitatively  different: While in \UP~one has only 5f electrons in orbitals with different degree of localization at the same sites as described by the dual model, in \CF~the subsystems of localized 4f and itinerant 3d electrons are spatially separated in different layers.\\
The anomalous behavior of CEF excitations in \CF~is a strong indirect evidence for the 3d spin exciton in superconducting FeAs layers. We note that subsequently their presence has been directly confirmed by INS in the isostructural \LF~compound that has no 4f electrons \cite{wakimoto:10}.

\section{Conclusion and Outlook}
\label{sec:conclusion}

In this survey we discussed the theoretical investigations on the collective spin exciton of strongly correlated f-electron systems. These modes are found inside small quasiparticle excitation gaps resulting from superconductivity, hidden order or directly from the hybridization. They are bound states of quasiparticles whose signature is a pole or sharp resonance in the dynamical magnetic response. Usually they are confined to the narrow vicinity of a nesting wave vector \bQ~  of the Fermi surface which frequently is a zone boundary vector. This leads to a large magnetic response already in the normal state and is favorable for satisfying the bound state condition. 

The spin exciton has the following signatures: i) The resonance energy $\omega_r$ fulfils $\omega_r/2\Delta<1$ where $2\Delta$ is the superconducting or normal state quasiparticle excitation gap. ii) If the latter shows the temperature dependence of an order parameter the energy and intensity of the resonance follows that temperature dependence. iii) The piling up of intensity at the resonance position leads to the evolution of a spin gap (intensity depletion) at lower energies.

For spin exciton modes to form in unconventional superconductors, it is necessary that the gap respects the sign change condition  $\Delta_{\bk+\bQ}=-\Delta_\bk$ at the resonance wave vector. Only then the non-vanishing coherence factors close to the gap threshold
allow for a singular magnetic response and bound state formation. This condition gives an important criterion for the symmetry of the SC gap functions and was used in all heavy fermion superconductors discussed above. In Kondo semiconductors or heavy electron metals when the gap is due to hybridization or hidden or magnetic order, no such condition is necessary or possible because the dynamic magnetic response function is always of the semiconducting type without coherence factors that might vanish. Therefore the observation of a
resonance in this case does not allow direct conclusions on the symmetry of the hidden order.

Of particular interest is the behavior of the spin exciton in a magnetic field. As a triplet mode it should split into three components as predicted for cuprates. Such pioneering experiments were carried out in \CC~with the surprising result of a splitting into a doublet or only a shifting of the resonance, depending on field direction. This was explained as an effect of anisotropies of g-factors and quasiparticle interactions. It is not clear what happens to the splitting at larger field when the upper mode merges into the quasiparticle continuum and  the lower mode approaches softening conditions. Therefore INS investigation in higher fields than previously are desirable in this compound.

The resonance and spin gap formation may also have indirect consequences on thermal transport and  microwave conductivity  because it influences the lifetime of charge carriers. In the superconductors this leads to strong deviations from canonical BCS type temperature behavior of these quantities. Such effects have possibly been found in cuprates and pnictides that exhibit spin resonances  but have received little theoretical attention so far.
Finally we mention that the spin exciton modes may have been found in other non-superconducting f electron systems like the second 4f hexaboride SmB$_6$ \cite{fuhrman:15} and the cage compound CeFe$_2$Al$_{10}$. On the other hand experiments to find a spin exciton in the unconventional heavy fermion superconductor UBe$_{13}$ \cite{hiess:14} are inconclusive.

\section*{Acknowledgments}
The authors sincerely thank Ilya Eremin for collaboration and for preparing some of the figures used in this article.

\bibliographystyle{ws-procs9x6} 
\bibliography{arXiv}

\end{document}